\documentclass{iopart}
\usepackage{iopams,cite,graphicx,tabularx}  
\begin{document}

% User defined commands
\newcommand{\rd}{\mathrm{d}}
\newcommand{\expct}[1]{\langle #1 \rangle}
\newcommand{\Expct}[1]{\left\langle #1 \right\rangle}
\newcommand{\cum}[1]{{\langle #1 \rangle}_{\rm c}}
\newcommand{\Cum}[1]{{\left\langle #1 \right\rangle}_{\rm c}}
\newcommand{\diff}[2]{\frac{\mathrm{d} #1}{\mathrm{d} #2}}
\newcommand{\prt}[2]{\frac{\partial #1}{\partial #2}}
\renewcommand{\(}{\left(}
\renewcommand{\)}{\right)}
\renewcommand{\[}{\left[}
\renewcommand{\]}{\right]}
\newcommand{\im}{\mathrm{Im}}
\newcommand{\re}{\mathrm{Re}}
\newcommand{\Std}{\mathrm{Std}}
\newcommand{\cdf}{\mathrm{cdf}}
\newcommand{\unit}[1]{~\mathrm{#1}}
\newcommand{\ep}{\varepsilon}
\newcommand{\eqref}[1]{(\ref{#1})}
\newcommand{\chem}[1]{$\mathrm{#1}$}
\renewcommand{\thefootnote}{\arabic{footnote}}

\title[Experimental approaches to universal out-of-equilibrium scaling laws]{Experimental approaches to universal out-of-equilibrium scaling laws:\\ turbulent liquid crystal and other developments}

\author{Kazumasa A Takeuchi}

\address{Department of Physics, the University of Tokyo,
 7-3-1 Hongo, Bunkyo-ku, Tokyo 113-0033, Japan.}
\ead{kat@kaztake.org}
\begin{abstract}
This is a brief survey of recent experimental studies
 on out-of-equilibrium scaling laws,
 focusing on two prominent situations where
% non-trivial fixed points in the sense of renormalization group
 non-trivial universality classes
 have been identified theoretically:
 absorbing-state phase transitions and growing interfaces.
First the article summarizes main results obtained
 for electrically-driven turbulent liquid crystal,
 which exhibited the scaling laws for the directed percolation class
 at the transition between two turbulent regimes,
 and those for the Kardar-Parisi-Zhang class in the supercritical phase
 where one turbulent regime invades the other.
Other experimental investigations
 on these universality classes and related situations
 are then overviewed and discussed.
Some remarks on analyses of these scaling laws
 are also given from the practical viewpoints.
\end{abstract}

%\pacs{}
%\submitto{J. Stat. Mech.}
%\maketitle

%\vspace{20pt}

\section{Introduction}  \label{sec:1}

Physics of critical phenomena and other scale-invariant problems
 like the Brownian motion is a landmark in statistical mechanics,
 which has provided us with simple and unified views on
 diverse phenomena in nature
 \cite{Goldenfeld-Book1992,Henkel-Book1999}.
Scale invariance refers to absence of characteristic scales.
It is known to arise, typically,
 when a system is undergoing a second-order phase transition.
A prototypical example is the Ising model.
While patches or islands of up/down spins
 with finite characteristic lengths $\xi$ are formed
 in the ferro- and paramagnetic phases,
 $\xi$ diverges at the critical temperature,
 leading to fractal spin configuration
 where fluctuations take place in all scales, so that scale invariance arises.
Descriptions of such systems should then also be scale-invariant,
 wiping out most microscopic details like the choice of material
 or the mechanism of interactions
 as one moves toward larger scales of interest.
Roughly, this is how universality arises in such scale-invariant phenomena,
 often in terms of power laws
 characterized by universal scaling (critical) exponents
 -- the idea formulated and established notably by Wilson
 by means of renormalization group theory \cite{Wilson-RMP1975}.
These scaling laws usually depend solely on global properties of the systems,
 such as spatial dimensions, symmetry, and conservation laws,
 which group various systems into a few universality classes
 that share the same critical exponents and scaling functions.
This is now deeply understood for equilibrium systems
 by theoretical frameworks like renormalization group and
 conformal field theory \cite{Goldenfeld-Book1992,Henkel-Book1999},
 and, importantly, firmly underpinned by ample experimental evidence:
 for example, the Ising-class criticality has been observed in real experiments
 of magnets, liquid-vapor systems and alloys,
 to name but a few \cite{Henkel-Book1999}.

It is then natural to expect similar universality
 for systems driven out of equilibrium,
 since the concept does not seem to necessitate being at equilibrium.
Indeed, theoretical developments have adapted renormalization group
 and field-theoretic techniques to out-of-equilibrium systems,
 identifying a number of universality classes
 that have no counterpart in equilibrium, hence truly out of equilibrium
 \cite{Tauber-Book2013,Hinrichsen-AP2000,Henkel.etal-Book2009,Barabasi.Stanley-Book1995}.
Two noteworthy classes in this context are the directed percolation (DP) class,
 governing continuous phase transitions into an absorbing state
 \cite{Tauber-Book2013,Hinrichsen-AP2000,Henkel.etal-Book2009},
 and the Kardar-Parisi-Zhang (KPZ) class for generic nonlinear growth processes
 \cite{Tauber-Book2013,Kardar.etal-PRL1986,Barabasi.Stanley-Book1995,Meakin-PR1993,HalpinHealy.Zhang-PR1995,Krug-AP1997}\footnote{
Note that the KPZ problem can be mapped to
 an equilibrium model under disordered environment,
 namely the directed polymer in random medium
 \cite{Barabasi.Stanley-Book1995,Meakin-PR1993,HalpinHealy.Zhang-PR1995,Krug-AP1997}.
Moreover, in one dimension, the steady state of the KPZ class satisfies
 a certain form of the fluctuation-dissipation theorem 
 \cite{Kardar.etal-PRL1986,Barabasi.Stanley-Book1995,Meakin-PR1993,HalpinHealy.Zhang-PR1995,Krug-AP1997}.
%A recent study also showed that dynamic correlators
% of one-dimensional unharmonic oscillator chains can be described
% by steady-state properties of the KPZ class \cite{Mendl.Spohn-a2013}.
These facts, however, do not indicate that the KPZ class
 describes equilibrium systems in the usual way universality classes do,
 so are not contradictory to the out-of-equilibrium nature of the KPZ class
 or corresponding growth processes.
}, both corresponding to the simplest case
 without extra symmetry or conservation laws.
These classes are now theoretically well understood,
 with ample numerical evidence for universality
 in a wide variety of models and situations.
Nevertheless, they had remained quite elusive in real experiments,
 as pointed out repeatedly in the literature (e.g., \cite{Hinrichsen-AP2000,Henkel.etal-Book2009,Barabasi.Stanley-Book1995})
 as a serious issue to be settled and understood.

In the author's view, main difficulties
 in studying such out-of-equilibrium scaling laws experimentally
 stem from a chief characteristic of out-of-equilibrium systems:
 formation of dissipative structure.
For instance, the Rayleigh-B\'enard convection forms convective rolls
 and other types of structure, which behave as single dynamical units
 when they start to oscillate or fluctuate
 \cite{Ciliberto.Bigazzi-PRL1988,Daviaud.etal-PRA1990}.
The effective system size is then given by the aspect ratio of the system,
 typically in the order of tens or a hundred at most for fluid systems,
 which is much smaller than the Avogadro number.
This makes truly macroscopic behavior difficult to access.
Similarly, since each degree of freedom has a relatively large scale,
 thermal noise is in most cases negligible.
Stochastic processes then result from other sources of effective noise,
 such as chaos or external heterogeneity,
 which however tend to introduce complex and possibly long-range
 noise correlations, as well as quenched disorder.
It is well known that these factors can affect or even destroy
 the scaling laws that one would expect in their absence
 \cite{Hinrichsen-AP2000,Henkel.etal-Book2009,Barabasi.Stanley-Book1995}.
In addition, the presence of ordered structure can easily introduce
 effectively long-range interactions to the system,
 which may also have a decisive impact on the scaling laws.

To overcome all these difficulties, in a series of papers
 \cite{Takeuchi.etal-PRL2007,Takeuchi.etal-PRE2009,Takeuchi.Sano-PRL2010,Takeuchi.etal-SR2011,Takeuchi.Sano-JSP2012}
 the author and his coauthors
 have focused on the electroconvection of nematic liquid crystal
 \cite{deGennes.Prost-Book1995,Kai.Zimmermann-PTPS1989}.
This is a system much studied in the context of pattern formation
 and is argued to be analogous to the Rayleigh-B\'enard convection,
 except that liquid crystal molecules are driven by an electric field
 through the Carr-Helfrich instability \cite{deGennes.Prost-Book1995}.
Thanks to its efficient driving mechanism, the electroconvection
 is usually studied in a very thin container of liquid crystal,
 which implies a very large system size.
Moreover, working with turbulent regimes
 called the dynamic scattering modes 1 and 2 (DSM1 and DSM2)
 \cite{deGennes.Prost-Book1995,Kai.Zimmermann-PTPS1989},
% or, more precisely,
% regimes of spatiotemporal chaos where correlation decays exponentially
% in both space and time,
 which are actually regimes of spatiotemporal chaos
 with exponentially decaying correlation in space and time,
 one can safely eliminate
 the effect of long-range correlations.
Finally, the source of noise in this system
 is intrinsic turbulent fluctuations,
 which overwhelm uncontrolled heterogeneities of the experimental setup.
In this simple yet non-trivial system
 for studying out-of-equilibrium scaling laws,
 the author and coworkers
 have investigated a phase transition between the DSM1 and DSM2 states
 \cite{Takeuchi.etal-PRL2007,Takeuchi.etal-PRE2009}
 as well as growth processes in the DSM2 phase,
 where stable DSM2 clusters invade the metastable DSM1 phase
 \cite{Takeuchi.Sano-PRL2010,Takeuchi.etal-SR2011,Takeuchi.Sano-JSP2012}.

The article is organized as follows.
First, section~\ref{sec:2} summarizes main results obtained
 for the DSM1-DSM2 transition and for the DSM2 growth processes.
As will be explained there,
 the former turned out to exhibit the DP-class criticality
 \cite{Takeuchi.etal-PRL2007,Takeuchi.etal-PRE2009},
 while the latter shows the scaling laws of the KPZ class
 \cite{Takeuchi.Sano-PRL2010,Takeuchi.etal-SR2011,Takeuchi.Sano-JSP2012}.
Section~\ref{sec:3} overviews other experimental studies
 related to these two problems, with more weight on recent active developments.
In the hope of facilitating these directions of research,
 section~\ref{sec:4} describes some general remarks and practical hints
 on analyses of these scaling laws,
 followed by concluding remarks in section~\ref{sec:5}.
Note, however, that this article is \textit{not} intended to be
 a comprehensive review of these matters:
 there are many experimental studies that
 the author was not able to cover in this article,
 especially on the case where they are related in some indirect manner,
 or they fit better to different theoretical contexts
 (e.g., cases with long-range interactions, quenched disorder, etc.).
The author also reminds the readers that the article deals with
 scaling laws associated with absorbing-state transitions and growth processes,
 leaving aside other interesting out-of-equilibrium scaling laws,
 in particular those for fully developed turbulence \cite{Frisch-Book1995}.

\section{Main results of the turbulent liquid crystal experiment}
\label{sec:2}

\begin{figure}[t]
 \centering
  \includegraphics[clip, width=\hsize]{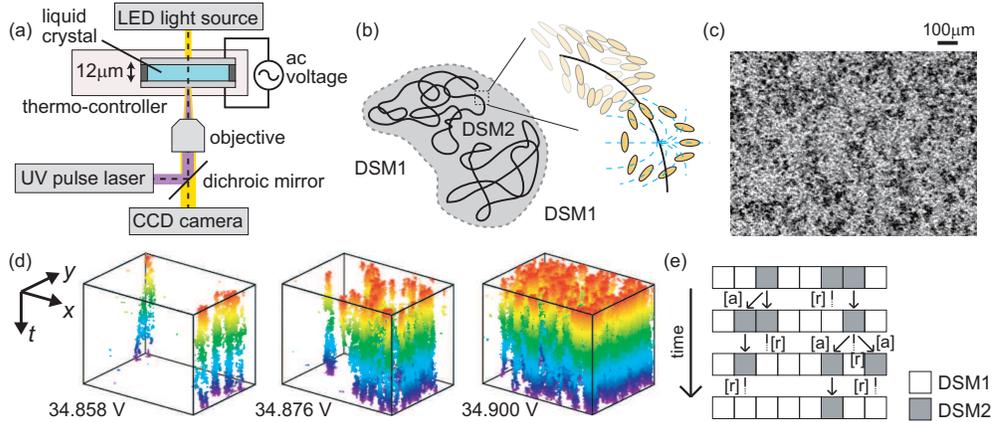}
  \caption{Experimental setup for the electroconvection and spatiotemporal intermittency. (a) Schematic illustration of the experimental setup. Note that this is a much simplified sketch, omitting many components such as filters, lenses, and isothermal chambers. For full descriptions, see \cite{Takeuchi.etal-PRE2009} for the experiments presented in section \ref{sec:2.1} and \cite{Takeuchi.Sano-JSP2012} for those in section \ref{sec:2.2}. (b) Sketch of the DSM2 state, composed of many disclinations, or line defects, in the liquid-crystal orientation. Blue dashed curves are contour lines of equal orientation. (c) Spatiotemporal intermittency at a voltage slightly above $V_{\rm c}$, namely 35.153~V (see also video 1). Active DSM2 patches (black) coexist amid the absorbing DSM1 region (white). Shown here is an image after background removal and contrast enhancement, while raw images can be found in figure~1(b) in \cite{Takeuchi.etal-PRE2009} and movie~S1 in \cite{Takeuchi.etal-PRL2007}. (d) Spatiotemporal evolution of the DSM2 patches for three constant voltages near $V_{\rm c}$, shown in the range of $1206 \times 899$~$\mu$m$^2$ in space (the whole observation area) and $6.6$~s in time. (e) Contact process as a coarse-grained description of the observed spatiotemporal intermittency. Active (DSM2) sites activate neighbors [a] and/or relax to the inactive (DSM1) state [r], at constant rates determined by the applied voltage. Note that the minimum size of a DSM2 domain is known to be about $d/\sqrt{2}$ with the cell thickness $d$ \cite{Kai.etal-JPSJ1989} (here $d = 12\unit{\mu{}m}$), which roughly gives a length scale of each site in such lattice models. Panels (b,c,d,e) are reprinted from \cite{Takeuchi.etal-PRE2009} with adaptations.}
  \label{fig1}
\end{figure}%

This section describes main experimental results
 obtained for the liquid-crystal turbulence.
Specifically, section \ref{sec:2.1} deals with the DP-class critical behavior
 found at the DSM1-DSM2 transition
 \cite{Takeuchi.etal-PRL2007,Takeuchi.etal-PRE2009},
 and section \ref{sec:2.2} shows
 the KPZ-class interface fluctuations identified in the DSM2 growth processes
 \cite{Takeuchi.Sano-PRL2010,Takeuchi.etal-SR2011,Takeuchi.Sano-JSP2012}.
For complete descriptions,
 the readers are referred to the original papers,
 in particular \cite{Takeuchi.etal-PRE2009} and \cite{Takeuchi.Sano-JSP2012}.

The standard experimental setup for both studies
% (with some adjustments)
 is briefly outlined in figure~\ref{fig1}(a).
A thin container of inner dimensions 16~mm $\times$ 16~mm $\times$ 12~$\mu$m
% was constructed from glass plates with transparent electrodes
% and thin polyester films,
 was filled with nematic liquid crystal
 $N$-(4-methoxybenzylidene)-4-butylaniline (MBBA)
 doped with tetra-$n$-butylammonium bromide.
The inner surfaces of the container were treated so that
 liquid crystal molecules are aligned in a desired direction on the surfaces,
 here along an axis parallel to the surfaces, unless otherwise specified.
% either parallel or perpendicular
% (called planar or homeotropic alignment, respectively).
One can apply an alternating electric field perpendicularly to the sample,
 using transparent electrodes coated on the surfaces.
Here, a low frequency of the field was chosen, specifically 250~Hz,
 in order to work within the conductive regime of the electroconvection
 \cite{deGennes.Prost-Book1995,Kai.Zimmermann-PTPS1989}.
The convection was observed through the transmitted light,
 emitted from light-emitting diodes
 and captured by a charge-coupled device camera.
The temperature of the sample was kept constant with high precision
 throughout each experiment, by means of a thermocontroller
 and heat-insulating walls (the latter not shown in figure~\ref{fig1}(a)).
Lastly, in the presented setup, one can shoot ultraviolet laser pulses
 to the liquid crystal sample,
 which are used to trigger local nucleation of the DSM2 state
 in the DSM1 region \cite{Takeuchi.etal-PRE2009}.

In the low-frequency region chosen here, 
 one can observe a variety of patterns such as rolls and grids
 as the amplitude $V$ of the applied voltage is increased,
 finally reaching the regimes of spatiotemporal chaos, DSM1 and DSM2
 \cite{deGennes.Prost-Book1995,Kai.Zimmermann-PTPS1989},
 separated by the critical voltage $V_{\rm c}$.
The difference between DSM1 and DSM2 lies in their density
 of topological defects, or specifically, disclinations,
 in the liquid-crystal orientation (figure~\ref{fig1}(b)).
The DSM2 state consists of a high density of disclinations
 \cite{Kai.etal-JPSJ1989,Kai.etal-PRL1990},
 which leads, in macroscopic scales, to a lower light transmittance
 and to the loss of nematic anisotropy.
Spontaneous nucleation of the DSM2 state is therefore prevented
 by a high energy barrier for formation of topological defects.
While for $V \gg V_{\rm c}$ such events do occur in realistic time scales
 \cite{Kai.etal-JPSJ1989,Kai.etal-PRL1990},
 followed by domain growth of the stable DSM2 phase,
 for voltages near the threshold, spontaneous nucleation of the DSM2 state
 becomes a practically unobservable event.
The dynamics is then characterized by spatiotemporal intermittency
 of the DSM2 state, where dark DSM2 patches randomly migrate
 amid the bright DSM1 region (figure~\ref{fig1}(c) and video~1),
 sometimes splitting, sometimes disappearing.
This stochastic process
% can be regarded in the mesoscopic scales as a result of
 results from random transport, stretching, and shrinkage of disclinations,
 due to local chaotic flow of the electroconvection.
Note also that, for the DSM2 state to be formed and maintained,
 constituent disclinations should break surface anchoring
 of the liquid-crystal alignment \cite{Fazio.Komitov-EL1999}
 by reaching both the top and bottom surfaces of the container.
Therefore, despite the quasi-two-dimensional geometry of the container,
 the DSM1-DSM2 configuration is defined
 in a purely two-dimensional space.
This also implies that the minimum size of a single DSM2 domain
 is in the order of the cell thickness $d$,
 specifically $d/\sqrt{2}$ \cite{Kai.etal-JPSJ1989},
 which would give a cutoff length scale for coarse-grained descriptions
 of the DSM1-DSM2 dynamics at stake.

\subsection{Directed percolation class governing the DSM1-DSM2 transition \cite{Takeuchi.etal-PRL2007,Takeuchi.etal-PRE2009}}
\label{sec:2.1}

Here we overview the spatiotemporal intermittency described above,
 which characterizes the DSM1-DSM2 transition.
The two states can be distinguished by the light intensity (see video~1),
 but for an automatic binarization one should also take into account
 that the DSM1-DSM2 configuration evolves more slowly
 than local intensity fluctuations
 and that DSM2 domains cannot be smaller than $d^2/2$ \cite{Kai.etal-JPSJ1989}
 (see \cite{Takeuchi.etal-PRE2009} for details).
Figure~\ref{fig1}(d) shows results of such binarization,
 indicating how DSM2 patches evolve in space and time
 in the steady state,
 at some voltages $V$ slightly larger than $V_{\rm c}$.
This shows that, for larger $V$, more space is occupied by DSM2.
In contrast, for $V$ lower than the threshold,
 all DSM2 patches eventually disappear.
Therefore, one can choose the DSM2 area fraction, denoted by $\rho$,
 as an order parameter for this transition.

\begin{figure}[t]
 \centering
  \includegraphics[clip, width=0.9\hsize]{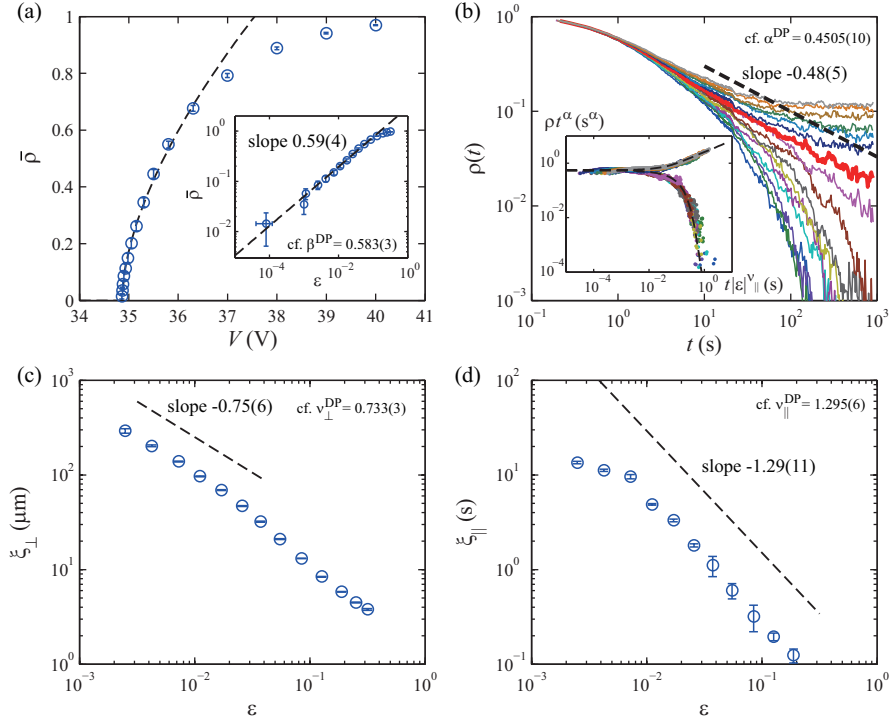}
  \caption{Critical behavior at the DSM1-DSM2 transition. (a) Time-averaged order parameter $\bar\rho$ (DSM2 area fraction) versus the applied voltage $V$. The inset shows the same data in logarithmic scales as a function of $\ep \equiv (V^2 - V_{\rm c}^2) / V_{\rm c}^2$. (b) Relaxation of the order parameter after quenching. Different colors correspond to different target voltages $V$, increasing from $34.86\unit{V}$ (bottom left) to $35.16\unit{V}$ (top right), with the critical case highlighted by the bold red line. The inset shows the same data with rescaled axes $t|\ep|^{\nu_\parallel}$ and $\rho t^\alpha$, which collapse onto the universal scaling function $F_\rho^{\rm DP}(\cdot)$ obtained numerically (dashed line). The upper and lower branches correspond to $V > V_{\rm c}$ and $V < V_{\rm c}$, respectively. (c,d) Correlation length $\xi_\perp$ (c) and correlation time $\xi_\parallel$ (d) in the steady state. The lengths measured in the direction of the initial molecular alignment are shown in (c), while the same exponent (within the error) was found in the perpendicular direction as well\cite{Takeuchi.etal-PRE2009} (see table~\ref{tbl1}). In all panels except the inset of (b), the dashed lines are guides for the eyes showing the estimated critical exponents. All panels are reprinted from \cite{Takeuchi.etal-PRE2009} with adaptations.}
  \label{fig2}
\end{figure}%

Figure~\ref{fig2}(a) shows the time-averaged order parameter $\bar\rho$
 against the applied voltage $V$ in the steady state,
 clearly indicating that the transition is continuous.
We therefore assume the usual scaling form
 $\bar\rho \sim (V - V_{\rm c})^\beta$,
% which is rewritten here as
 or, instead, an asymptotically equivalent one
\begin{equation}
 \bar\rho \sim (V^2 - V_{\rm c}^2)^\beta
\end{equation}
 following convention for the electroconvection,
 which takes into account its driving force
 proportional to $V^2$ \cite{deGennes.Prost-Book1995,Kai.Zimmermann-PTPS1989}.
The data indeed showed a clear power law over 3 digits,
 providing a precise estimate of the critical voltage
 at $V_{\rm c} = 34.856(4)\unit{V}$
 and the critical exponent $\beta = 0.59(4)$,
 as shown in the inset of figure~\ref{fig2}(a).
% plotted against $\ep \equiv (V^2 - V_{\rm c}^2) / V_{\rm c}^2$.
Here, the numbers in the parentheses
 indicate the range of errors in the last digit.
The estimate for $\beta$ therefore rules out, e.g.,
 the Landau mean-field exponent $1/2$,
 but instead agrees with the exponent for the DP class in $2+1$ dimensions,
 $\beta^{\rm DP} = 0.583(3)$
 \cite{Grassberger.Zhang-PA1996,Voigt.Ziff-PRE1997,Hinrichsen-AP2000,Henkel.etal-Book2009}.
Similarly,
% to characterize correlation, it is useful to measure
% the distribution of DSM1 intervals, 
 the correlation length $\xi_\perp$ and time $\xi_\parallel$ can be
 measured from DSM1 interval distributions in the steady state
 \cite{Takeuchi.etal-PRE2009}.
The length $l$ or the duration $\tau$ separating two neighboring DSM2 domains
 shows power-law distribution at criticality,
 $\mathrm{pdf}(l) \sim l^{-\mu_\perp}$
 and $\mathrm{pdf}(\tau) \sim l^{-\mu_\parallel}$,
 while for $V > V_{\rm c}$ it is cut off by an exponential tail as
 $\mathrm{pdf}(l) \sim \e^{-l / \xi_\perp}$
 and $\mathrm{pdf}(\tau) \sim \e^{-l / \xi_\parallel}$
 for large $l$ and $\tau$.
From this, $\xi_\perp$ and $\xi_\parallel$ were found to show
 power-law divergence at the critical point
 (figures~\ref{fig2}(c,d)) as follows:
\begin{equation}
 \xi_\perp \sim (V^2 - V_{\rm c}^2)^{-\nu_\perp}, \quad
 \xi_\parallel \sim (V^2 - V_{\rm c}^2)^{-\nu_\parallel}.
\end{equation}
The data yielded $\nu_\perp = 0.75(6)$ and $\nu_\parallel = 1.29(11)$,
 both in agreement with the DP-class values
 $\nu_\perp^{\rm DP} = 0.733(3)$ and $\nu_\parallel^{\rm DP} = 1.295(6)$.
Moreover, the values of the exponents $\mu_\perp$ and $\mu_\parallel$
 for the power-law distributions also supported the DP-class scenario,
 $\mu_\perp = 1.08(18)$ and $\mu_\parallel = 1.60(5)$
 \cite{Takeuchi.etal-PRE2009}, to be compared with
 $\mu_\perp^{\rm DP} = 1.204(2)$ and $\mu_\parallel^{\rm DP} = 1.5495(10)$.
This also confirms the following scaling relations:
\begin{equation}
 \mu_\perp = 2 - \beta / \nu_\perp,\quad
 \mu_\parallel = 2 - \beta / \nu_\parallel.  \label{eq:ScalingRelationMu}
\end{equation}

One can also study dynamic critical behavior by, e.g.,
 suddenly decreasing the applied voltage from a value deep in the DSM2 phase
 to a target value near the critical point.
Figure~\ref{fig2}(b) shows relaxation of the order parameter $\rho$
 after such critical quenching, which indicates power-law decay
 $\rho(t) \sim t^{-\alpha}$ with $\alpha = 0.48(5)$ at criticality
 (red bold curve)\footnote{
The critical voltage in the quench experiment was determined
 from figure~\ref{fig2}(b), independently
 of the estimate from the steady-state experiment.
This is to avoid influence of a slight, uncontrolled shift 
 in the experimental conditions
 during the days that separated the two sets of experiments
 \cite{Takeuchi.etal-PRE2009}.
}.
Theoretically, the standard scaling ansatz reads
 \cite{Hinrichsen-AP2000,Henkel.etal-Book2009}
\begin{equation}
 \rho(t) \sim t^{-\alpha} F_\rho(\ep t^{1/\nu_\parallel}),\quad \alpha = \beta/\nu_\parallel,  \label{eq:quench}
\end{equation}
 with $\ep \equiv (V^2 - V_{\rm c}^2) / V_{\rm c}^2$
 and a universal scaling function $F_\rho(\cdot)$.
This gives $\alpha^{\rm DP} = 0.4505(10)$ and
 correctly accounts for the experimental result.
Moreover, the scaling form in equation~\eqref{eq:quench} implies that
 data for different voltages should collapse
 on the single universal function $F_\rho(\cdot)$,
 when $\rho(t)t^\alpha$ is plotted against $\ep t^{1/\nu_\parallel}$.
This is indeed the case as shown in the inset of figure~\ref{fig2}(b),
 where the data in the main panel overlap onto the DP-class
 universal scaling function $F_\rho^{\rm DP}(\cdot)$
 obtained independently by a numerical simulation.
The result here substantiates, therefore,
 both the scaling relation $\alpha = \beta/\nu_\parallel$
 and the universality in the scaling function $F_\rho(\cdot)$.

It is also worthwhile to study another aspect of dynamic critical behavior,
 by shooting laser pulses
 to create a single DSM2 nucleus in the fully DSM1 state
 and track the resultant DSM2 cluster.
Power laws at criticality were then identified
 for the probability that the cluster survives until time $t$,
 as well as for the mean volume and the mean squared radius of the cluster,
 with the corresponding exponents of the DP universality class
 \cite{Takeuchi.etal-PRE2009}.
Actually, that survival probability is known to serve
 as another order parameter
% characterizing
 for the DP-class and related phase transitions,
 whose exponent $\beta'$ is generally different from $\beta$.
For the DP class, however,
 a special time-reversal symmetry called the rapidity symmetry
 leads to $\beta' = \beta$ \cite{Hinrichsen-AP2000,Henkel.etal-Book2009}.
The liquid-crystal experiment therefore underpins
 the presence of this nontrivial symmetry in a real system.
In such ways, the three sets of experiments
 for the steady state, critical quench, and critical spreading
 have revealed a total of 12 critical exponents,
 5 scaling functions, and 8 scaling relations, in full agreement with
 those characterizing the $(2+1)$-dimensional DP class
 \cite{Takeuchi.etal-PRL2007,Takeuchi.etal-PRE2009}.

Theoretically, the appearance of the DP-class criticality in this experiment
 can be accounted for on the basis of Pomeau's general argument
 on spatiotemporal intermittency \cite{Pomeau-PD1986}
 and Janssen and Grassberger's DP conjecture \cite{Janssen-ZPB1981,Grassberger-ZPB1982,Hinrichsen-AP2000,Henkel.etal-Book2009}.
Pomeau's argument \cite{Pomeau-PD1986} is illustrated in figure~\ref{fig1}(e):
 consider, for the sake of simplicity, a two-state lattice model
 with active or inactive local states,
 which correspond to the DSM2 and DSM1 states in the experiment, respectively.
An active site can activate neighbors and/or relax into the inactive state,
 both stochastically at constant rates.
This reproduces, at least qualitatively,
 spatiotemporal intermittency like the one observed
 in the liquid-crystal experiment (figure~\ref{fig1}(c) and video~1)
 and is nothing but a percolation process directed
 in the time forward direction.
Indeed, the model explained here, called the contact process,
 is recognized as a prototypical model exhibiting the DP-class transition
 \cite{Hinrichsen-AP2000,Henkel.etal-Book2009}.
% as repeatedly confirmed by numerical studies.
Of course it is an oversimplified model for the electroconvection,
 but then the DP conjecture implies that such differences do not change
 the critical behavior.
Specifically, the DSM1-DSM2 transition can be regarded as a transition
 into an absorbing state, i.e.,
 a global state that the system stays in forever once it enters,
 since, in practice,
 the DSM1 state admits no spontaneous nucleation of the DSM2 state.
The DP conjecture then states, in short, that
 phase transitions into a single absorbing state
 generically belong to the DP universality class, unless the system has
 additional symmetry, a conservation law, long-range interactions,
 or quenched disorder \cite{Janssen-ZPB1981,Grassberger-ZPB1982,Hinrichsen-AP2000,Henkel.etal-Book2009}.
This is true of the DSM1-DSM2 transition as argued in section~\ref{sec:1}
 and discussed in more detail in \cite{Takeuchi.etal-PRE2009}.
As a result, the DP-class criticality arises
 in such a real, microscopically complex system as well.

\subsection{Kardar-Parisi-Zhang class for the DSM2 growing interfaces \cite{Takeuchi.Sano-PRL2010,Takeuchi.etal-SR2011,Takeuchi.Sano-JSP2012}}
\label{sec:2.2}

\begin{figure}[t]
 \centering
  \includegraphics[clip, width=\hsize]{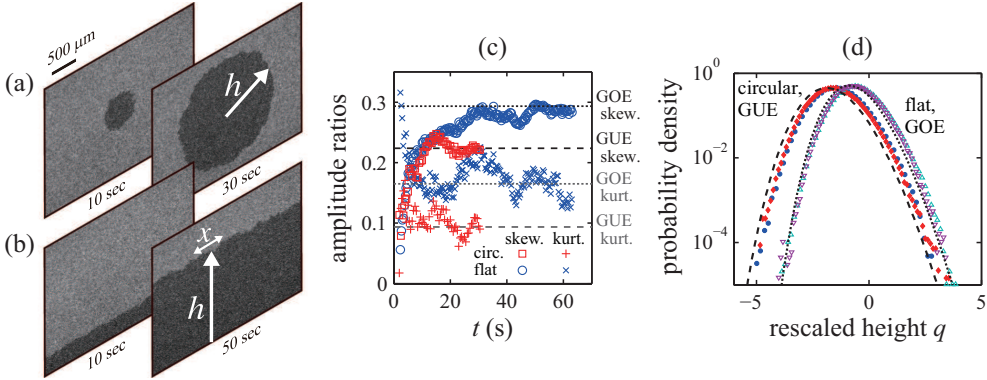}
  \caption{Growing DSM2 interfaces and one-point distribution. (a,b) Snapshots of a circular (a) and flat (b) interface. Elapsed time after emission of laser pulses is indicated. (c) Skewness $\expct{h^3}_{\rm c} / \expct{h^2}_{\rm c}^{3/2}$ and kurtosis $\expct{h^4}_{\rm c} / \expct{h^2}_{\rm c}^2$ (symbols), compared with the values for the GUE and GOE Tracy-Widom distributions (dashed and dotted lines). (d) One-point distribution of the rescaled local height $q \equiv (h - v_\infty t)/(\Gamma t)^{1/3}$ (symbols), compared with the GUE and GOE Tracy-Widom distributions (dashed and dotted lines). The data for the circular interfaces (solid symbols) were measured at $t = 10\unit{s}$ (blue \fullcircle) and $30\unit{s}$ (red \fulldiamond), while those for the flat interfaces were taken at $t = 20\unit{s}$ (turquoise \opentriangle) and $60\unit{s}$ (purple \opentriangledown). Note that the standard random variable for the GOE Tracy-Widom distribution is multiplied by $2^{-2/3}$ here, as suggested by analytical studies. All panels are reprinted from \cite{Takeuchi.Sano-JSP2012} with adaptations, with kind permission from Springer Science+Business Media.}
  \label{fig3}
\end{figure}

So far we have reviewed critical behavior at the DSM1-DSM2 transition,
 where a single cluster of the DSM2 state does not
 grow or shrink in an obvious way.
This is because, in terms of Pomeau's scenario in figure~\ref{fig1}(e),
 the activation rate is essentially balanced with the recession rate.
By contrast, when the applied voltage is sufficiently raised,
 the activation rate overwhelms the recession rate,
 leading to sustained growth of the DSM2 cluster
 bordered by a fluctuating interface (figures~\ref{fig3}(a,b)).
Microscopically, while disclinations inside the DSM2 cluster are maintained
 at a more or less constant density, on the border,
 strong chaotic flow continuously elongates, multiplies, and randomly transport
 disclinations, leading to the random local expansion
 of the DSM2 cluster in macroscopic scales.
One can easily observe such growing clusters
 using spontaneous nucleations at high applied voltages,
 regardless of the chosen alignment of liquid crystal molecules.
Under planar alignment, however, elliptic clusters are produced
 because of the anisotropy in the horizontal plane
 \cite{Kai.etal-JPSJ1989,Kai.etal-PRL1990}.
% because of different growth speeds in the directions
% parallel and perpendicular to the alignment.
A simple and controlled experiment can be performed, instead,
 by measuring isotropic clusters under homeotropic alignment,
%The simplest, isotropic growth is therefore realized
% under homeotropic alignment,
 where molecules are aligned perpendicularly to the surfaces,
 and by triggering nucleation by laser pulses
%Further, by using laser pulses
 instead of relying on spontaneous nucleation.
The use of laser also allows changing the cluster shape.
% one can control the moment of the nucleation,
% as well as its geometry by shaping laser beam as desired.
Figures~\ref{fig3}(a,b) show growing DSM2 clusters
 produced in such a way by point- and line-shaped laser pulses,
 exhibiting circular and flat interfaces, respectively.
One can therefore investigate statistical properties
 of these growing interfaces,
 as well as consequences of the different geometries.

Figures~\ref{fig3}(a,b) indicate that
 the interfaces roughen in the course of time
 with fractal-like intricate structure.
This suggests scale invariance of the process,
 which may therefore be described by another set of universal scaling laws.
This turned out to be indeed the case
 \cite{Takeuchi.Sano-PRL2010,Takeuchi.etal-SR2011,Takeuchi.Sano-JSP2012},
 characterized by the KPZ universality class,
 known as the simplest generic class for such stochastic growth processes
 due to short-range interactions
 \cite{Kardar.etal-PRL1986,Barabasi.Stanley-Book1995,Meakin-PR1993,HalpinHealy.Zhang-PR1995,Krug-AP1997}.
To characterize the interface fluctuations,
 one defines the local height $h(x,t)$ as in figures~\ref{fig3}(a,b),
 as a function of the lateral coordinate $x$ and time $t$.
Roughness is then quantified by the standard deviation of $h(x,t)$,
 measured within a strip of lateral length $l$ and averaged
 over space and ensemble, denoted by $w(l,t)$.
For scale-invariant growth, this quantity exhibits
 the following power laws called the Family-Vicsek scaling
 \cite{Family.Vicsek-JPA1985}:
\begin{equation}
 %w(l,t) \sim t^\beta F_w(lt^{-1/z}) \sim \begin{cases} l^\alpha & (l \ll l_*), \\ t^\beta & (l \gg l_*), \end{cases}  \label{eq:FamilyVicsekWidth}
 w(l,t) \sim t^\beta F_w(lt^{-1/z}) \sim \cases{l^\alpha & for $l \ll l_*$, \\ t^\beta & for $l \gg l_*$,\\}  \label{eq:FamilyVicsekWidth}
\end{equation}
 with a scaling function $F_w$,
 scaling exponents $\alpha, \beta$, and $z \equiv \alpha/\beta$,
 and a crossover length $l_* \sim t^{1/z}$.
It was indeed verified for the DSM2 growth
 for both circular and flat interfaces\footnote{
However, for circular interfaces, in general,
 spatial average in the definition of $w(l,t)$ results in a slight bias
 in the estimate of $\beta$ from the Family-Vicsek scaling
 \cite{Takeuchi.Sano-JSP2012}.
One should use instead the standard deviation defined by ensemble average,
 $W(t) \equiv (\expct{h^2} - \expct{h}^2)^{1/2} \sim t^{\beta}$.
},
 with the exponent values known exactly
 for the one-dimensional KPZ-class interfaces,
 $\alpha = 1/2, \beta = 1/3$, and $z = 3/2$
 \cite{Kardar.etal-PRL1986,Barabasi.Stanley-Book1995,Meakin-PR1993,HalpinHealy.Zhang-PR1995,Krug-AP1997}.
In particular, this implies the following equation
 for the height evolution:
\begin{equation}
 h(x,t) \simeq v_\infty t + (\Gamma t)^{1/3}\chi(x',t),  \label{eq:Height}
\end{equation}
 with two constant parameters $v_\infty$ and $\Gamma$,
 a rescaled random variable $\chi(x',t)$
 and a rescaled coordinate $x' \propto xt^{-2/3}$.

Now we are ready to probe finer statistical properties
 of interface fluctuations, namely their distribution and correlation,
 through the rescaled variable $\chi$.
First, the skewness and the kurtosis were computed from the cumulants,
 defined by $\cum{h^3}/\cum{h^2}^{3/2}$ and $\cum{h^4}/\cum{h^2}^2$,
 respectively\footnote{
Definitions of the cumulants are given, up to the fourth order, by
 $\cum{h^2} \equiv \expct{\delta h^2}$,
 $\cum{h^3} \equiv \expct{\delta h^3}$, and
 $\cum{h^4} \equiv \expct{\delta h^4} - 3\expct{\delta h^2}^2$,
 with ensemble average $\expct{\cdots}$ and $\delta h \equiv h - \expct{h}$.
}.
The results in figure~\ref{fig3}(c) show that, first,
 the skewness and the kurtosis are non-zero,
 implying non-Gaussian statistics, and second,
 they are significantly different between the circular and flat interfaces
 (blue and red symbols, respectively).
In particular, the observed values were found to be quite close to those
 for apparently unrelated distributions developed in random matrix theory
 \cite{Mehta-Book2004},
 namely the largest-eigenvalue distribution of random matrices
 in the Gaussian unitary ensemble (GUE) for the circular interfaces
 (dashed line) and in the Gaussian orthogonal ensemble (GOE)
 for the flat interfaces (dotted line)\footnote{
The GUE and GOE random matrices and the corresponding Tracy-Widom distributions
 are defined as follows.
First, consider an $N \times N$ complex Hermitian matrix
 (hence $\bar{A}_{ij} = A_{ji}$) with random matrix elements.
Specifically, the real and imaginary parts are independently drawn
 from the Gaussian distribution, satisfying $\expct{A_{ij}}=0$ for all $i,j$,
 $\expct{A_{ii}^2} = N$,
 and $\expct{(\re A_{ij})^2} = \expct{(\im A_{ij})^2} = N/2$ for $i \neq j$.
This ensemble of random matrices is called GUE.
Similarly, GOE is defined by replacing complex Hermitian matrices
 with real symmetric matrices.
Then, whether GUE or GOE, the matrix $A$ has $N$ real random eigenvalues,
 the largest of which scales
 as $\lambda_{\rm max} \simeq 2N + N^{1/3}\chi$ for large $N$,
 in the normalization adopted here.
The distribution of this random variable $\chi$ in the limit $N\to\infty$
 is called the Tracy-Widom distribution.
Analytic expressions are available for GOE, GUE,
 and their symplectic counterpart
 \cite{Tracy.Widom-CMP1994,Tracy.Widom-CMP1996}.
In passing, it is suggestive to compare
 the above expression for $\lambda_{\rm max}$ and
 equation~\eqref{eq:Height} for the growing interfaces.
}.
This implies that the random variable $\chi$ asymptotically coincides,
 in the sense of distribution,
 with the variable $\chi_{\rm GUE}$ and $\chi_{\rm GOE}$ obeying
 these particular distributions, called
 the GUE and GOE Tracy-Widom distributions
 \cite{Tracy.Widom-CMP1994,Tracy.Widom-CMP1996}\footnote{
Note however that, for the sake of simplicity,
 $\chi_{\rm GOE}$ in this article is defined by
 multiplying $2^{-2/3}$ to the standard definition in random matrix theory,
 as suggested by analytical results for solvable growth models
 \cite{Prahofer.Spohn-PRL2000,Kriecherbauer.Krug-JPA2010,Sasamoto.Spohn-JSM2010,Corwin-RMTA2012}.
}.
This nontrivial hypothesis can be checked directly
 by determining the parameters $v_\infty$ and $\Gamma$
 in the way described in section~\ref{sec:4.2}
 (see \cite{Takeuchi.Sano-JSP2012} for details),
 and comparing histograms of the rescaled height
 $q \equiv (h-v_\infty t)/(\Gamma t)^{1/3}$
 with the theoretical distributions.
The experimental data were then indeed found very close
 to the Tracy-Widom distributions (figure~\ref{fig3}(d))
 without adjustable fitting parameters.
To add, the slight deviations apparent in figure~\ref{fig3}(d) turned out
 to be finite-time corrections in the cumulants \cite{Takeuchi.Sano-JSP2012},
 mainly in the mean.
They were found to decay
 as $\cum{q^n} - \cum{\chi_{\rm GUE/GOE}^n} \sim t^{-n/3}$,
 except for $n=2$ and 4 in the circular case, where the difference was
 too small to extract any systematic change in time.
In short, the fluctuations of the circular and flat interfaces
 for $t\to\infty$
 were shown to exhibit
% in the asymptotic limit
 the GUE and GOE Tracy-Widom distributions, respectively,
 at least up to the fourth-order cumulant.
The convergence to the different distributions
 is counterintuitive, since the circular interfaces become flatter and flatter
 as time elapses.
This is however not contradictory, if one recalls scale invariance
 in this system, for which memory from the initial condition
 would remain forever.
Note that the Tracy-Widom distributions in the KPZ class were actually
 first found by analytical studies for solvable models
 \cite{Johansson-CMP2000,Prahofer.Spohn-PRL2000}
 and recently even for the KPZ equation
 \cite{Sasamoto.Spohn-PRL2010,Amir.etal-CPAM2011,Calabrese.LeDoussal-PRL2011}\footnote{
See \cite{Kriecherbauer.Krug-JPA2010,Sasamoto.Spohn-JSM2010,Corwin-RMTA2012}
 for reviews on these analytical developments.},
 which is the paradigmatic continuum equation for the KPZ class
 \cite{Kardar.etal-PRL1986}:
\begin{equation}
 \prt{h}{t} = \nu\nabla^2 h + \frac{\lambda}{2}(\nabla h)^2 + \eta(x,t),  \label{eq:KPZeq}
\end{equation}
 with $\expct{\eta(x,t)}=0$ and
 $\expct{\eta(x,t)\eta(x',t')} = D\delta(x-x')\delta(t-t')$.
% after several steps of subtle and nontrivial mathematical mapping
% and computations.
These analytical results, together with the experimental support,
 imply that the Tracy-Widom distributions
 are universal characteristics of the $(1+1)$-dimensional KPZ class.
In particular, the single KPZ class splits
 into at least two universality subclasses
 separating the circular (or curved) and flat interfaces.

\begin{figure}[t]
 \centering
  \includegraphics[clip, width=\hsize]{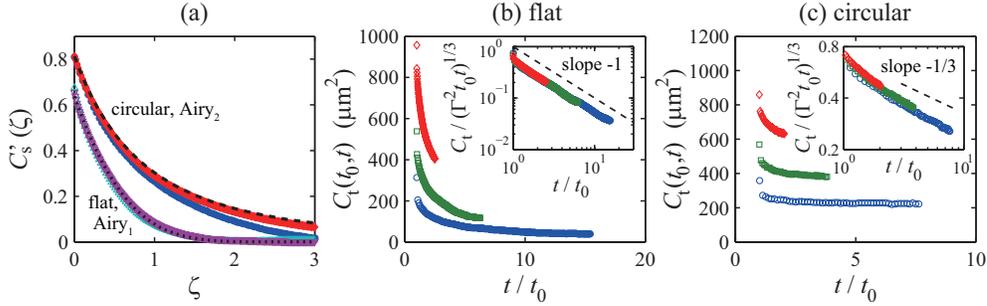}
  \caption{Spatial and temporal correlation in DSM2 interface fluctuations. (a) Rescaled spatial correlation function $C'_{\rm s}(\zeta;t) \equiv C_{\rm s}(l;t) / (\Gamma t)^{2/3}$ against rescaled length $\zeta \equiv (Al/2)(\Gamma t)^{-2/3}$ (symbols), compared with the Airy$_1$ and Airy$_2$ correlations (dashed and dotted lines, respectively). The same color code as in figure~\ref{fig3}(d) is used. (b,c) Time correlation function $C_{\rm t}(t,t_0)$ for the flat (b) and circular (c) interfaces, measured with different $t_0$: $t_0 = 4\unit{s}, 10\unit{s}, 25\unit{s}$ for (b) and $4\unit{s}, 8\unit{s}, 15\unit{s}$ for (c) from bottom to top. The rescaled correlation function $C_{\rm t}(t,t_0) / (\Gamma^2 t_0 t)^{1/3}$ is shown in the insets, with guides for the eyes showing the power of the asymptotic decay. All panels are reprinted from \cite{Takeuchi.Sano-JSP2012} with adaptations, with kind permission from Springer Science+Business Media.}
  \label{fig4}
\end{figure}%

Correlation can also be studied experimentally.
For the spatial correlation,
 analytical studies have shown that it is described by
 time correlation of the particular stochastic processes
 called the Airy$_2$ and Airy$_1$ processes,
 for the curved and flat interfaces, respectively
 \cite{Kriecherbauer.Krug-JPA2010,Sasamoto.Spohn-JSM2010,Corwin-RMTA2012}.
Specifically, the two-point spatial correlation
\begin{equation}
C_{\rm s}(l;t) \equiv \expct{h(x+l,t)h(x,t)} - \expct{h(x+l,t)}\expct{h(x,t)},  \label{eq:SpatialCorrFuncDef}
\end{equation}
 is given by
\begin{equation}
 C_{\rm s}(l;t) \simeq (\Gamma t)^{2/3} g_i\( \frac{Al}{2}(\Gamma t)^{-2/3} \)  \label{eq:AiryCorrelation}
\end{equation}
 with $g_i(\zeta) \equiv \expct{\mathcal{A}_i(t'+\zeta)\mathcal{A}_i(t')} - \expct{\mathcal{A}_i(t')}^2$, the Airy$_i$ process $\mathcal{A}_i(t')$
 ($i=1$ or $2$),
 and certain coefficient $A$ linked to parameters in the KPZ equation,
 given by $A=\sqrt{2\Gamma/v_\infty}$ in the case of isotropic growth
 \cite{Takeuchi.Sano-JSP2012}.
This result for solvable models was also reproduced experimentally
 as shown in figure~\ref{fig4}(a), again with finite-time corrections
 as expected.
This actually implies that the difference
 between the circular and flat interfaces
 is not necessarily merely quantitative,
 because it is theoretically known that the Airy$_2$ correlation
 for the circular interfaces decreases as $g_2(\zeta) \sim \zeta^{-2}$,
 while the Airy$_1$ correlation for the flat interfaces decays
 faster than exponentially \cite{Bornemann.etal-JSP2008}.

In contrast to the spatial correlation, time correlation remains
 intractable by analytical studies,
 leaving room for unique contributions by experiments
 and other empirical or phenomenological approaches.
The liquid-crystal experiment showed that it is also different
 between the two cases, qualitatively, as shown in figures~\ref{fig4}(b,c)
 for the two-time correlation function
\begin{equation}
C_{\rm t}(t,t_0) \equiv \expct{h(x,t)h(x,t_0)} - \expct{h(x,t)}\expct{h(x,t_0)}.  \label{eq:TemporalCorrFuncDef}
\end{equation}
For the flat interfaces (figure~\ref{fig4}(b)),
 correlation decays toward zero,
 following the natural scaling ansatz
 $C_{\rm t}(t,t_0) \simeq (\Gamma^2 t_0 t)^{1/3} F_{\rm t}(t/t_0)$
 that takes into account aging effect in the scale-invariant growth,
 with $F_{\rm t}(\tau) \sim \tau^{-1}$ (inset).
By contrast, the results for the circular interfaces (figure~\ref{fig4}(c))
 suggest that the same ansatz does not work as well
 (as rescaled data do not perfectly overlap)
 and, moreover, the correlation decays as $F_{\rm t}(\tau) \sim \tau^{-1/3}$
 in the rescaled unit, hence remains strictly positive in the original scale.
These claims are not definitive since $t/t_0$ was varied by less than a digit
 experimentally, but the same result was obtained numerically
 \cite{Takeuchi-JSM2012},
 and the latter claim is also supported by phenomenological theory
 due to Singha \cite{Singha-JSM2005}.
% with an explicit expression for $C_{\rm t}(t,t_0)$.
Similarly, the flat and circular interfaces were found to be different
 in many other quantities, notably in persistence properties,
 as summarized in table~1 of \cite{Takeuchi.Sano-JSP2012}.

According to analytical developments \cite{Prahofer.Spohn-PRL2000,Kriecherbauer.Krug-JPA2010,Sasamoto.Spohn-JSM2010,Corwin-RMTA2012},
 the curved (circular) and flat subclasses are not the only ones
 constituting the KPZ class.
Another important subclass is that for the stationary interfaces,
 which still grow but without changing their statistical properties.
To study this subclass theoretically, the initial condition is set
 to be a stationary interface itself,
 e.g., a flat interface in the limit $t \to\infty$,
 which is equivalent to the one-dimensional Brownian motion
 for the $(1+1)$-dimensional KPZ class
 \cite{Barabasi.Stanley-Book1995,Meakin-PR1993,HalpinHealy.Zhang-PR1995,Krug-AP1997}.
Then the height difference from the initial time was analytically shown
 to obey Baik and Rains' $F_0$ distribution
 \cite{Baik.Rains-JSP2000,Prahofer.Spohn-PRL2000}.
This subclass is, however, unable to be reached experimentally or numerically,
 unless a stationary interface is artificially prepared
 as an initial condition\footnote{\label{ft:Stat}
It is not clear, at least strictly,
 how the properties of the stationary subclass appear in the steady state
 reached within a finite-size system,
 since one cannot avoid influence from the boundary
 in the asymptotic limit.
% if one considers height difference
% between times $t_0$ and $t_0 + \Delta t$,
% the final state of finite-size interfaces is obtained
% by taking the $\Delta t\to\infty$ limit first,
% while for the stationary subclass one takes $t_0\to\infty$ and then
% $\Delta t\to\infty$.
As a matter of fact, no exact solution is obtained yet
 for systems with finite sizes.
}.
Instead, for the evolution between two finite times $t_0$ and $t_0 + \Delta t$,
 it was found both experimentally and numerically that
 the rescaled height difference undergoes crossover
 between the flat and stationary subclasses,
 described by universal scaling functions parametrized by $\Delta t/t_0$
 \cite{Takeuchi-PRL2013}.
To summarize, it is now clear that
 for the $(1+1)$-dimensional KPZ class,
% provides a unique situation where
 universal statistical properties out of equilibrium
 can be analyzed very deeply by analytical means, revealing
 their characteristic geometry- or initial-condition-dependence.
The DSM2 growth was found to be a useful experimental system,
 not only to test but complement these analytical developments.

\section{Other experimental systems}  \label{sec:3}

While the previous section dealt entirely with the liquid-crystal experiment,
 this is by no means an only experimental system
 to study absorbing-state transitions or growing interfaces.
On the contrary, universality implies, at least in principle,
 that the same macroscopic descriptions apply
 to a vast variety of phenomena and situations,
 as is indeed established on the numerical side.
The present section overviews other experimental systems
 related to absorbing-state transitions and to growing interfaces,
 especially those concerning the DP class and the KPZ class.

\subsection{Experiments on absorbing-state transitions} \label{sec:3.1}

\begin{table}
 %\begin{minipage}{\textwidth}
  \caption{Critical exponents for absorbing-state transitions in experiments.$^{\rm a}$}
  \label{tbl1}
  \catcode`?=\active \def?{\phantom{0}}
  %\begin{indented}
  %\item[]
  \begin{tabular}{@{}llllllll} \br
   (1+1)D System & $\beta$ & $\nu_\perp$ & $\nu_\parallel$ & $\mu_\perp$ & $\mu_\parallel$ \\ \mr
   annular Rayleigh-B\'enard \cite{Ciliberto.Bigazzi-PRL1988} & & $0.5$ & & $1.9(1)$ & $1.9$ \\
   annular Rayleigh-B\'enard \cite{Daviaud.etal-PRA1990} & & $0.5$ & $0.5$ & $1.7(1)$ & $2.0(1)$ \\
   linear Rayleigh-B\'enard \cite{Daviaud.etal-PRA1990} & $0.30(5)$ & $0.50(5)$ & $0.50(5)$ & $1.6(2)$ & $2.0(2)$ \\
   depinning in paper wetting \cite{Buldyrev.etal-PRA1992} & & \multicolumn{2}{l}{$\nu_\perp/\nu_\parallel = 0.63(4)$} & & \\
   viscous fingering \cite{Michalland.etal-EL1993} & $0.45(5)$ & & & $0.64(2)$ & $0.61(2)$ \\
   fluid vortices \cite{Willaime.etal-PRE1993} & $0.5$ & & & & $1.7$ \\
   Taylor-Dean \cite{Degen.etal-PRE1996} & $1.30(26)$ & $0.64, 0.53$ & $0.73$ & $1.67(14)$ & $1.74(16)$ \\
   Taylor-Couette \cite{Colovas.Andereck-PRE1997} & $1$ & $0.4$ & & $1.4$-$2.5$ & \\
   granular flow \cite{Daerr.Douady-N1999,Hinrichsen.etal-PRL1999} & & \multicolumn{2}{l}{$\nu_\parallel-\nu_\perp = 1$} & & \\
   torsional Couette \cite{Cros.LeGal-PF2002} & $0.30(1)$ & $0.53(5)$ & & &  \\
   ferrofluidic spikes \cite{Rupp.etal-PRE2003} & $0.30(5)$ & $1.1(2)$ & $0.62(14)$ & $1.70(5)$ & $2.1(1)$ \\
   lateral heat convection \cite{Lepiller.etal-PF2007} & $0.27(3)$ & $0.30(4)$ & $0.75(3)$ & & \\
   viscoelastic Taylor-Couette \cite{Latrache.etal-PRE2012} & $1.5(1)$ & $0.78(5)$ & $1.00(5)$ & & $1.31$ \\ \mr
%   pipe-flow turbulence \\ \mr
   DP \cite{Jensen-JPA1999} & $0.276$ & $1.097$ & $1.734$ & $1.748$ & $1.841$ \\
   & \multicolumn{2}{l}{$\nu_\perp/\nu_\parallel = 0.633,$} & \multicolumn{2}{l}{$\nu_\parallel-\nu_\perp = 0.637$} & \\ \br
   (2+1)D System & $\beta$ & $\nu_\perp$ & $\nu_\parallel$ & $\mu_\perp$ & $\mu_\parallel$ \\ \mr
   liquid columns \cite{Pirat.etal-PRL2005} & $0.56(5)$ & & & & \\
   DSM1-DSM2 \cite{Takeuchi.etal-PRL2007,Takeuchi.etal-PRE2009}$^{\rm b}$ & $0.59(4)$ & $0.75(6)$ & $1.29(11)$ & $1.08(18)$ & $1.60(5)$ \\
   & & $0.78(9)$ & & $1.19(12)$ \\ \mr
   DP \cite{Grassberger.Zhang-PA1996,Voigt.Ziff-PRE1997} & $0.583(3)$ & $0.733(3)$ & $1.295(6)$ & $1.204(2)$ & $1.5495(10)$ \\ \br
  \end{tabular}
  \noindent $^{\rm a}$ Number in parentheses is the range of error given by the authors of each article.\\
  \noindent $^{\rm b}$ The estimates for $\nu_\perp$ and $\mu_\perp$ were given for the two different axes in the plane, parallel (upper raw) and perpendicular (lower raw) to the initial molecular alignment.
  %\end{indented}
 %\end{minipage}
\end{table}

Table~\ref{tbl1} presents a list of experiments
 -- most probably not a complete list --
 performed in situations where one might expect
 the DP-class criticality to arise.
This shows that the values of experimentally measured critical exponents
 are rather widely distributed instead of being universal,
 as already discussed in section~\ref{sec:1},
 in a strong contrast with the status of numerical investigations
 \cite{Hinrichsen-AP2000,Henkel.etal-Book2009}.
One notices in particular that,
 even if some of the exponents agree with the DP-class values,
 others may be significantly different
 (see, e.g., the values for the linear Rayleigh-B\'enard convection
 \cite{Daviaud.etal-PRA1990} or for the ferrofluidic spikes
 \cite{Rupp.etal-PRE2003} in table~\ref{tbl1}).
Therefore, for experimental investigations on absorbing-state transitions,
 it is essential to determine a complete set
 of independent critical exponents,
 such as $(\beta, \nu_\perp, \nu_\parallel)$
 or $(\beta, \mu_\perp, \mu_\parallel)$ for the DP class\footnote[1]{
Although one needs in general four independent exponents
 to characterize absorbing-state transitions, 
 for the DP class the number reduces to three thanks to the rapidity symmetry
 $\beta = \beta'$, as briefly discussed in section~\ref{sec:2.1}.
},
 and to measure as many exponents as possible in a given setup.
This would clarify whether a given experimental system belongs
 to the DP class or another,
 helping to make further steps to elucidate
 what are essential reasons for being or not being in the DP class
 experimentally.
In the following, brief descriptions are given
 for some interesting experimental systems
 studied in relation to absorbing-state transitions.

\subsubsection{Spatiotemporal intermittency of ferrofluidic spikes.}

One of the most thorough experiments
 on one-dimensional spatiotemporal intermittency was carried out
 by Rupp \etal \cite{Rupp.etal-PRE2003} using magnetic fluids.
When a horizontal layer of magnetic fluid
 is subjected to a strong enough vertical magnetic field,
 the Rosensweig instability leads to formation of liquid spikes.
Using a sharp edge of a cylindrical electromagnet
 to produce an inhomogeneous magnetic field,
 Rupp \etal trapped ferrofluidic spikes on this edge,
 which were regularly aligned at a constant interval.
Then they added an alternating component to the magnetic field,
 which induced spatiotemporal chaos or spatiotemporal intermittency
 of the spikes, depending on the amplitude of the alternating component.
They analyzed the steady state of this spatiotemporal intermittency
 for different amplitudes of the alternating component,
 determining five critical exponents,
 $\beta, \nu_\perp, \nu_\parallel, \mu_\perp, \mu_\parallel$,
 as listed in table~\ref{tbl1}.
Interestingly, while those for the order parameter ($\beta$)
 and for the spatial correlation ($\nu_\perp$ and $\mu_\perp$)
 indicated the values for the $(1+1)$-dimensional DP class,
 those for the temporal correlation ($\nu_\parallel$ and $\mu_\parallel$)
 were found significantly different from the DP class\footnote{
The authors of \cite{Rupp.etal-PRE2003}
 cited an erroneous value of $\mu_\parallel$ for the DP class,
 which led them to conclude
 that four of the five exponents agreed with the DP class.
}.
%Rupp \etal mentioned, as potential reasons for this discrepancy,
% the possible presence of a continuum of states
% between the absorbing and active states, as well as the possibility
% for spontaneous nucleation from the absorbing state.
A critical-quench experiment could also be performed for this system
 along the procedure described in section~\ref{sec:2.1},
 which would 
%To
 further characterize the dynamic aspect of the transition
% it would be interesting to perform a critical-quench experiment
% along the procedure described in section~\ref{sec:2.1},
% in order to determine
 through the exponent $\alpha$ ($= \beta/\nu_\parallel$
 if the scaling ansatz \eqref{eq:quench} holds)
 and the scaling function $F_\rho(\cdot)$.

\subsubsection{Onset of pipe-flow turbulence.}

The onset of a sustained turbulent state
 in a flow along a long pipe is an outstanding issue in fluid mechanics,
 which dates back to the seminal work by Reynolds \cite{Reynolds-PTRSL1883}
 in 1883.
% and has recently marked a considerable progress
% in relation to absorbing-state transitions.
Since the laminar flow in this setup, called the Hagen-Poiseuille flow,
 is known to be linearly stable for all Reynolds numbers,
 the problem consists in persistence of a turbulent state
 induced by a finite disturbance to the system.
For relatively low Reynolds numbers $\mathrm{Re} \lesssim 2700$,
 this turbulent state was shown to have a rather localized structure
 along the stream, hence called the turbulent puff
 \cite{Wygnanski.Champagne-JFM1973,Hof.etal-S2004}.
Recently, Hof and coworkers have performed a series of experiments
 \cite{Hof.etal-S2004,Hof.etal-N2006,Hof.etal-PRL2008,Avila.etal-S2011}
 using carefully designed pipes of unprecedented lengths
 (typically $15\unit{m}$ and maximum $30\unit{m}$,
 which amount to 3750 and 7500 times the pipe diameter, respectively).
They found that the survival probability of a turbulent puff decays
 exponentially in time, suggesting a memoryless stochastic decay process,
 with the mean life time that grows very rapidly with $\mathrm{Re}$
 but does \textit{not} diverge near the apparent critical point
 \cite{Hof.etal-N2006,Hof.etal-PRL2008}.
At first glance, this result may imply
 the absence of a well-defined critical Reynolds number.
However, their subsequent work \cite{Avila.etal-S2011} showed that
 a puff can also split into two as a result of another stochastic event,
 whose typical time grows rapidly with decreasing $\mathrm{Re}$.
These results were further confirmed by another experimental group
 \cite{Kuik.etal-JFM2010}
 as well as by numerical simulations
 \cite{Hof.etal-N2006,Avila.etal-JFM2010,Avila.etal-S2011}.
In view of all these results,
 Hof's group located the critical point at $\mathrm{Re} = 2040 \pm 10$,
 at which the decaying rate and the splitting rate are balanced
 \cite{Avila.etal-S2011}\footnote{
Strictly, the critical point is not exactly given by the balance
 of the two local transition rates.
However, for the pipe-flow turbulence, 
 both the life time and the splitting time grow superexponentially fast
 near the critical point
 \cite{Hof.etal-PRL2008,Avila.etal-JFM2010,Avila.etal-S2011},
 so that the difference from the true critical point
 would be smaller than the given range of error \cite{Avila.etal-S2011}.
}

Apart from determining the critical point, Hof \etal's scenario
 based on the decaying and splitting of turbulent puffs
 is reminiscent of Pomeau's argument \cite{Pomeau-PD1986}
 and the contact process (figure~\ref{fig1}(e)).
This leads to the simplest scenario where
 the critical dynamics of the pipe-flow turbulence is governed
 by the DP universality class, if one puts aside
 the rarity of the DP-class criticality in experiments
 (table~\ref{tbl1}).
% which seemingly exhibit similar local dynamics.
Therefore, a conclusion should be made after
 direct analysis of its critical behavior,
 but the time scale needed for a single decay or splitting event
 at the critical point is desperately long to do so
 by experiments or by direct numerical simulations.
A way to circumvent this difficulty is studying
 whether common factors that affect the DP class are present or not
 in pipe-flow turbulence.
As an example of this direction of research, Hof's group
 measured interval distributions between two turbulent puffs
 both experimentally \cite{Samanta.etal-JFM2011}
 and numerically \cite{Avila.Hof-PRE2013}
 and found no indication of long-range correlation,
 though they were unable to reach the steady state near the critical point,
 where one would expect power-law distribution of laminar intervals
 characterized by the exponent $\mu_\perp$.
It could also be informative to study quenching dynamics
 after a sudden decrease of the Reynolds number
 \cite{Samanta.etal-JFM2011,Moxey.Barkley-PNAS2010}
 in the context of the critical quench experiment
 for absorbing-state transitions (see section~\ref{sec:2.1}).
%After this series of progress
% in the understanding of the pipe-flow turbulence,
In a different line,
 Hof \etal's approach for measuring the life time and the splitting time
 seems to provide a useful way to revisit other hydrodynamic systems
 exhibiting spatiotemporal intermittency.
% such as the plane Couette flow \cite{Shi.etal-PRL2013}.
Hof's group indeed applied it to the plane Couette flow successfully
 \cite{Shi.etal-PRL2013}.
This system is particularly interesting,
 because they found, numerically, power-law distributions
 of laminar intervals near the critical point \cite{Shi.etal-PRL2013}.

\subsubsection{Reversible-irreversible transitions in colloidal suspensions.}

Another interesting and nontrivial example of absorbing-state transitions
 was found in transitions between reversible and irreversible motions
 of colloids under a periodic shear.
Pine \etal \cite{Pine.etal-N2005} studied colloidal motion
 in a viscous fluid confined in concentric cylinders (a Couette cell),
 the inner one being rotated back and forth slowly.
In the absence of colloids, such a flow at low Reynolds numbers
 is described by the Stokes equation, and hence reversible
 with respect to a periodic displacement of the boundary.
This is also true of the system mixed with non-Brownian colloids,
 i.e., particles large enough to neglect their Brownian motion,
 when they are shared periodically
 at small shear strain amplitudes; however, Pine \etal found that
 the particle motion becomes irreversible and effectively diffusive
 when the strain amplitude $\gamma$ exceeds
 a certain threshold $\gamma_{\rm c}$ \cite{Pine.etal-N2005}.
The origin of this irreversible motion was argued to be in chaos
 due to hydrodynamic interactions of nearby particles, 
 but the connection between this microscopic chaos
 and the macroscopic irreversibility remained to be shown.

\begin{figure}[t]
 \centering
  \includegraphics[clip, width=.9\hsize]{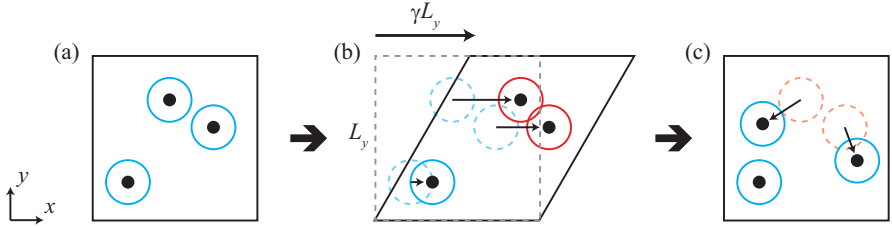}
  \caption{Sketch of the model proposed by Cort\'e \etal \cite{Corte.etal-NP2008}. (a) Initial configuration. The range of particle interactions is shown by blue circles. (b) A shear of strain amplitude $\gamma$ replaces each particle by $\gamma y$, causing collisions of nearby particles (red). (c) After each period of oscillatory shear, particle collisions result in random displacement of the collided particles.}
  \label{fig5}
\end{figure}%

A reasonable explanation for this reversible-irreversible transition was later
 given by Cort\'e \etal \cite{Corte.etal-NP2008} by means of a simple model
 (figure~\ref{fig5}).
Consider a two-dimensional space and
 assume a finite interaction range of particles.
These particles are distributed randomly at the initial time.
When the system is sheared along the $x$ axis
 with the strain amplitude $\gamma$,
 each particle is displaced by $\gamma y$ (figure~\ref{fig5}(b)).
This may cause nearby particles to encounter each other,
 which results in random displacement of the particle positions
 after a cycle of the periodic shear,
 while those which did not experience collision return
 to their original positions (figure~\ref{fig5}(c)).
Then, for small $\gamma$, the particles eventually find such a configuration
 that no pair of particles collides any more,
 whereas such a state is never reached for large enough $\gamma$.
This is the reversible-irreversible transition.
It can be regarded as a transition to an absorbing state,
 since a collisionless configuration
 implies no further rearrangement of particles.
Here, however, the transition is coupled with a conserved field
 because of the fixed particle number.
Therefore, one expects another universality class to arise,
 called the conserved DP class or the Manna class
 \cite{Hinrichsen-AP2000,Henkel.etal-Book2009,Lubeck-IJMPB2004},
 instead of the DP class.
Indeed, Cort\'e \etal's model as well as the colloidal experiment showed
 indications of an an absorbing-state transition,
 though they did not obtain a quantitative agreement
 with the conserved DP class \cite{Corte.etal-NP2008}
 ($\beta = 0.45(2)$ and $\nu_\parallel = 1.33(2)$
 from two-dimensional simulations \cite{Corte.etal-NP2008},
 to be compared with $\beta = 0.639(9)$ and $\nu_\parallel = 1.225(29)$
 for the conserved DP class \cite{Henkel.etal-Book2009,Lubeck-IJMPB2004}; 
 $\beta = 0.45(10)$ and $\nu_\parallel = 1.1(3)$ for the experiment
 in three dimensions \cite{Corte.etal-NP2008},
 to be compared with $\beta = 0.840(12)$ and $\nu_\parallel = 1.081(27)$
 for the conserved DP class \cite{Henkel.etal-Book2009,Lubeck-IJMPB2004}).
Moreover, Cort\'e \etal reported an interesting finding that
 a rheological quantity, namely the elastic component of the complex modulus,
 $\eta''$, behaves similarly to the order parameter $\rho$
 \cite{Corte.etal-NP2008},
 though the reason for this is yet to be clarified.
After this work, Menon and Ramaswamy \cite{Menon.Ramaswamy-PRE2009}
 introduced a lattice version of Cort\'e \etal's model
 and found a quantitative agreement with the conserved DP class
 both for two and three dimensions.
They also argued that long-range hydrodynamic interactions
 may explain the different critical exponents observed in the experiment. 

Some interesting studies have been carried out subsequently.
Franceschini \etal \cite{Franceschini.etal-PRL2011}
 used rods instead of spherical particles,
 and revealed two different transitions as $\gamma$ was increased,
 the first one for the formation of nematic alignment
 and the second one being the reversible-irreversible transition.
Interestingly, in this setup, the latter transition was nicely accounted for
 by the conserved DP class, with the correct values
 of $\beta$ and $\nu_\parallel$ for three dimensions
 \cite{Franceschini.etal-PRL2011}.
From different perspectives,
 Mangan, Reichhardt, and Olson Reichhardt argued
 an analogous transition
% can be observed
 in driven systems with quenched disorder,
 such as vortices in superconductors (or other types of particles)
 under periodic driving \cite{Mangan.etal-PRL2008}
 and plastic depinning of trapped particles
 \cite{Reichhardt.Reichhardt-PRL2009},
 on the basis of their numerical simulations.
Indeed, Okuma \etal \cite{Okuma.etal-PRB2011,Okuma.etal-JPSJ2012}
 experimentally identified the predicted transitions
% the reversible-irreversible transition
 using vortices driven by a Corbino disk made of an amorphous superconductor
 (\textit{a}-)Mo$_x$Ge$_{1-x}$.
In their experiments,
 the reversible-irreversible transition under periodic driving
 gave $\nu_\parallel = 1.3(3)$
 and the depinning transition under constant driving yielded
 $\nu_\parallel = 1.26(15)$.
Finally, it would be worth mentioning
 that the existence of the conserved DP class
 was recently challenged by numerical studies due to Basu \etal
 \cite{Basu.etal-PRL2012,Basu.etal-EPJB2013}.
They claimed that common models for the conserved DP class
 actually showed the DP-class exponents,
 when they used a ``natural'' initial condition that takes into account
 correlated configuration of inactive particles in the stationary state,
 hence the previously reported exponents for the conserved-DP class
 were numerical artifacts.
This is still controversial as no analytic argument is given yet,
 especially on the different field-theoretic descriptions of the two classes
 \cite{Hinrichsen-AP2000,Henkel.etal-Book2009,Lubeck-IJMPB2004,Vespignani.etal-PRL1998,Bonachela.Munoz-PA2007}.
In any case, though, it is true that the two universality classes have
 very similar sets of the critical exponents as well as the scaling functions
 \cite{Lubeck-IJMPB2004},
 usually indistinguishable in experiments.
In this context, studying effect of walls,
 originally proposed for numerical simulations
 \cite{Bonachela.Munoz-PA2007}, may also be useful for experiments.

\subsection{Experiments on growing interfaces}  \label{sec:3.2}

Now we turn our attention to interface fluctuations
 due to local growth processes.
Because of their ubiquity and importance,
 a great number of experiments have been carried out
 for a wide variety of systems, such as fluid flow in porous media,
 paper wetting, growing bacterial colony, to name but a few
 \cite{Barabasi.Stanley-Book1995,Meakin-PR1993,HalpinHealy.Zhang-PR1995,Krug-AP1997,Alava.etal-AP2004}.
%As mentioned in section~\ref{sec:1}, 
However, these studies have reported
 a wide range of the exponent values, mostly significantly different
 from those for the KPZ class;
 for one-dimensional interfaces,
 the KPZ-class values are $\alpha = 1/2$ and $\beta = 1/3$,
 while experiments tend to indicate larger values,
 typically $0.6 \lesssim \alpha \lesssim 0.8$ and $\beta \approx 0.6$
\cite{Barabasi.Stanley-Book1995,Meakin-PR1993,HalpinHealy.Zhang-PR1995,Krug-AP1997} (see, e.g., table~11.1 in \cite{Barabasi.Stanley-Book1995}).
For some of the experiments, the deviation from the KPZ class
 can be understood as consequences
 of quenched disorder \cite{Barabasi.Stanley-Book1995,Tang.Leschhorn-PRA1992,Buldyrev.etal-PRA1992,Amaral.etal-PRL1994,Amaral.etal-PRE1995,Tang.etal-PRL1995,Leschhorn.etal-APB1997,Csahok.etal-JPA1993,Csahok.etal-PA1993},
 long-range correlation \cite{Medina.etal-PRA1989},
 and/or power-law noise distribution \cite{Zhang-JP1990,Horvath.etal-PRL1991},
 which are common factors that intervene in typical experimental systems.
%Specifically, one notices that a good number of experiments
% for one-dimensional interfaces indicated,
% roughly, $0.6 \lesssim \alpha \lesssim 0.8$ and $\beta \approx 0.6$
% \cite{Barabasi.Stanley-Book1995,Meakin-PR1993,HalpinHealy.Zhang-PR1995,Krug-AP1997} (see, e.g., table~11.1 in \cite{Barabasi.Stanley-Book1995}),
% which are the values expected under quenched noise
% \cite{Csahok.etal-JPA1993,Csahok.etal-PA1993}
% or power-law noise distribution
%% with reasonable values of the exponent
%% with exponent near $-4$
% \cite{Zhang-JP1990,Horvath.etal-PRL1991}.
In particular, imbibition in porous media has been studied
 in depth along these lines, for which
 the presence of such effects and consequences in scaling behavior
 are now rather clear in theory and experiments \cite{Alava.etal-AP2004}.

Among those factors affecting the scaling laws,
 effect of quenched disorder has been particularly investigated
 and seems to be relevant in many experiments.
First of all, in the presence of quenched noise,
 interfaces may be pinned or unpinned, according to the driving strength.
While scaling behavior is expected to remain unchanged
 when interfaces are driven strongly,
 a few different universality classes may arise
 near the depinning transition,
 depending on the anisotropy and the nonlinearity
 of the growth process at stake \cite{Tang.etal-PRL1995}.
% in a different set of universal scaling laws,
% sometimes called the quenched KPZ class.
If the nonlinearity has its origin in kinematics
 and hence vanishes at the transition,
 as expected for simple isotropic growth processes,
 the critical dynamics will be governed
 by the quenched Edwards-Wilkinson class
 \cite{Tang.etal-PRL1995,Amaral.etal-PRL1994,Leschhorn.etal-APB1997}\footnote{
The Edwards-Wilkinson equation refers to the linear continuum equation
 obtained by setting $\lambda=0$ in equation \eqref{eq:KPZeq}.
If one replaces the usual noise term $\eta(x,t)$ by quenched one,
 $\eta(x,h)$, one obtaines the quenched Edwards-Wilkinson equation.
One can similarly define the quenched KPZ equation.
}.
By constrast, if the nonlinearity remains
 because of, e.g., anisotropy in the growth process,
 the quenched KPZ class will take over.
For this case, and for one-dimensional interfaces
 growing along the anisotropy axis,
% Tang and Leschhorn \cite{Tang.Leschhorn-PRA1992}
% and Buldyrev \etal \cite{Buldyrev.etal-PRA1992}
% proposed lattice models, independently, which allowed them to describe
 there exist lattice models that allow us to describe
 the depinning transition in terms of critical DP clusters
 \cite{Tang.Leschhorn-PRA1992,Buldyrev.etal-PRA1992,Amaral.etal-PRE1995}.
This leads to the following value of the roughness exponent $\alpha$
 for the interfaces pinned at criticality:
 $\alpha = \nu_\perp^{\rm DP} / \nu_\parallel^{\rm DP} \approx 0.633$.
% \cite{Tang.Leschhorn-PRA1992,Buldyrev.etal-PRA1992,Amaral.etal-PRE1995}.
For the dynamic exponent $z = \alpha/\beta$,
% one needs to consider depinned interfaces, but near the critical point,
 scaling arguments yield $z = 1$, hence $\alpha = \beta \approx 0.633$
 \cite{Tang.Leschhorn-PRA1992,Amaral.etal-PRE1995,Tang.etal-PRL1995}.
%One should however be aware that these exponents describe asymptotic dynamics
% only when the system is infinitesimally close to the critical point.
%When interfaces are growing, these qKPZ values should be valid
% \cite{Tang.Leschhorn-PRA1992}.
However, if the growth direction is tilted with respect to the anisotropy axis,
 another set of exponents is expected \cite{Tang.etal-PRL1995}.
Moreover, one should be aware that these exponents are valid
 only up to finite crossover length and time scales,
 unless the interfaces are exactly at the depinning transition,
 hence not moving.
From somewhat different approaches,
% Csah\'ok \etal \cite{Csahok.etal-JPA1993,Csahok.etal-PA1993} performed
 dimension analysis for the quenched KPZ equation
 yields $\alpha = 3/4$ and $\beta = 3/5$ for one dimension
 \cite{Csahok.etal-JPA1993,Csahok.etal-PA1993},
 again up to finite crossover scales.
Experimental values of $\alpha$ and $\beta$ tend to be close
 to these few sets of the scaling exponents,
% expected under quenched noise,
 though similar values can also be obtained
 from noise with long-range correlation \cite{Medina.etal-PRA1989}
 or from power-law noise distribution \cite{Zhang-JP1990,Horvath.etal-PRL1991}.

On the other hand, there have been reported a few experiments
 showing the KPZ-class exponents, the number growing rather rapidly
 in recent years.
These experiments are overviewed in the following subsections,
 apart from the liquid-crystal experiment already described
 in section~\ref{sec:2.2}.
Note that there are also a few indirect realizations
 of the KPZ-class scaling laws,
 specifically, in rupture lines \cite{Kertesz.etal-F1993,Engoy.etal-PRL1994}
 and in crystal facets \cite{Degawa.etal-PRL2006},
 which are not dealt with in this article.

\subsubsection{Colony growth of bacteria and other cells.}

Growth of bacterial colonies on agar is one of the systems well studied
 in the context of pattern formation \cite{Matsushita.etal-B2004},
 showing a variety of patterns like dendric and concentric ones.
One can also observe compact colonies for certain concentrations
 of agar and nutrients, which can therefore be regarded as realizations
 of growing interfaces.
Vicsek \etal \cite{Vicsek.etal-PA1990} studied such interfaces
 using \textit{Escherichia coli} and \textit{Bacillus subtilis}
 and reported $\alpha = 0.78(7)$ and $\alpha \approx 0.74$, respectively.
This latter result was later reproduced
 by Wakita \etal \cite{Wakita.etal-JPSJ1997}
 with a similar exponent value $\alpha \approx 0.78$.
However, Wakita \etal also used a mutant strain of \textit{B. subtilis}
 that does not secrete surfactant,
 in order to study another type of compact growing colonies
 exhibiting less rough interfaces.
It is in this case that their data were found to be consistent 
 with the KPZ-class exponents $\alpha = 1/2$ and $\beta = 1/3$
 \cite{Wakita.etal-JPSJ1997}\footnote{
Wakita \etal did not provide an estimate of $\beta$, but they presented
 data for the roughness growth \cite{Wakita.etal-JPSJ1997},
 which are consistent with $\beta = 1/3$
 albeit the narrow range of the scaling region.
}.
Later, Hallatschek \etal \cite{Hallatschek.etal-PNAS2007}
 developed an interesting experiment on colony formation,
 using two strains of bacteria with different fluorescent proteins
 but otherwise genetically identical.
The two strains segregate as the colony expands,
 forming superdiffusive boundaries in between.
Hallatschek \etal explained this result
 by assuming the KPZ-class exponents for the edge of the colony,
 though they did not measure these exponents from the experimental data
 \cite{Hallatschek.etal-PNAS2007}.

Recently, Huergo \etal performed analogous experiments for eukaryotic cells,
 namely Vero cells \cite{Huergo.etal-PRE2010,Huergo.etal-PRE2011}
 and cancer cells HeLa \cite{Huergo.etal-PRE2012},
 which form colonies directly on a Petri dish.
Then they found the two KPZ-class exponents $\alpha$ and $\beta$
 for all the above cases, which may be surprising
 in view of the usually non-KPZ behavior of the bacterial colonies
 and of more complex characters of eukaryotic cells.
Moreover, for Vero cells, Huergo \etal realized
 both flat and circular interfaces,
 which are particularly intriguing from the viewpoint
 of the recent developments
 on the geometry-dependent universality in the one-dimensional KPZ class
 (see section~\ref{sec:2.2}).
Therefore, performing analyses developed for the liquid-crystal turbulence
 \cite{Takeuchi.Sano-JSP2012} would be an important attempt
 for testing the robustness of this particular universality
 in such very different experimental systems.

\subsubsection{Slow combustion of paper.}

Another interesting experimental example of the KPZ class
 was reported on slow, flameless combustion of paper,
 almost concurrently with Wakita \etal's work on bacteria.
While an earlier experiment on slow combustion indicated $\alpha = 0.71(5)$
 \cite{Zhang.etal-PA1992},
 Timonen's group found the KPZ-class exponents
 by a series of highly controlled experiments
 \cite{Maunuksela.etal-PRL1997,Myllys.etal-PRE2001}:
 they prepared paper with uniform distribution of oxidizer KNO$_3$,
% which promotes combustion,
 and realized uniform heat transfer, by controlling air flow
 and by compensating heat loss at paper boundaries
 by means of heater filaments.
% using heater filaments that follow the combustion front.
In addition to the KPZ-class exponents,
 they also determined values of the coefficients in the KPZ equation
 \cite{Myllys.etal-PRE2001,Maunuksela.etal-EPJB2003}
 and amplitude ratios in the saturated regime \cite{Myllys.etal-PRE2001},
 i.e., the long-time regime where the interface roughness
 does not grow any more because of a finite size.
Then they found, in terms of the quantity
 called the universal coupling constant, $g^* = 0.79(9), 0.76(8), 1.0(2)$
 for three different sets of samples \cite{Myllys.etal-PRE2001}.
This quantity is now understood to be
 $g^* = \cum{\chi_0^2}^{-3/2} \approx 0.81$, with the random variable $\chi_0$
 of the Baik-Rains $F_0$ distribution for the stationary interfaces
 \cite{Prahofer.Spohn-PRL2000}.
Their results therefore indicate that the saturated regime
 of finite-size systems may also be characterized
 by the Baik-Rains $F_0$ distribution,
 though this statement should be confirmed by measuring other cumulants
 from the experimental and numerical sides,
 as well as by analytical studies on finite-size systems
 (see footnote~\ref{ft:Stat} on page~\pageref{ft:Stat}).

In subsequent work, Timonen and coworkers measured first-passage properties
 of interface fluctuations, 
 namely the persistence probability \cite{Merikoski.etal-PRL2003},
 i.e., the probability that fluctuations remain positive (or negative)
 over given durations in time (or lengths in space).
Their results for the saturated regime showed
 theoretically predicted power laws both in space and time.
In contrast, for the growing regime (flat interfaces),
 their temporal persistence probability did not show
 asymmetry between positive and negative fluctuations
 \cite{Merikoski.etal-PRL2003},
 as opposed to earlier numerical work \cite{Kallabis.Krug-EL1999}
 and to the liquid-crystal experiment \cite{Takeuchi.Sano-JSP2012}.
Similarly, the spatial persistence probability was found to decay algebraically
 for the paper-combustion experiment, in contrast to an exponential decay
 found in the liquid-crystal experiment \cite{Takeuchi.Sano-JSP2012}
 which was later supported by an analytical study
 \cite{Ferrari.Frings-JSM2013}.
This apparent discrepancy may be due to a rather narrow time window
 for the growing regime in the paper-combustion experiment,
 as the authors discussed \cite{Merikoski.etal-PRL2003},
 but this needs to be clarified by further investigations.

Finally, Timonen's group also studied the one-point distribution.
They found that the experimentally obtained histograms
 for the growing and saturated regimes were reasonably fitted
 by the GOE Tracy-Widom and Baik-Rains $F_0$ distributions, respectively,
 by adjusting the mean and the variance of these theoretical distributions
 to the experimental data \cite{Miettinen.etal-EPJB2005}.
However, when the mean and the variance are adjusted as free parameters,
 or, equivalently, normalized to zero and unity, respectively,
 the difference between these distributions becomes barely visible
 only below $10^{-4}$
 (see figure~\ref{fig6} and discussions in section~\ref{sec:4.2}),
 which is obscured by statistical and experimental errors
 in the data presented in \cite{Miettinen.etal-EPJB2005}.
They also provided values of the skewness
 for the growing and saturated regimes, specifically 0.33 and 0.32,
 respectively, which however lay between
 the skewness of the GOE Tracy-Widom distribution, 0.2935,
 and that of the Baik-Rains $F_0$ distribution, 0.359,
 without favoring one of them.
To avoid the loss of information by normalizing the distributions,
 one can measure the two constant parameters $v_\infty$ and $\Gamma$
 in equation~\eqref{eq:Height} and rescale the height variable by
 $q \equiv (h - v_\infty t)/(\Gamma t)^{1/3}$ (section~\ref{sec:4.2}).
This can be compared with the theoretical distributions
 without the need of \textit{ad hoc} fitting,
 which are readily distinguishable in this case.
%These observations imply that further study is called for to resolve
% the distribution functions in the paper-combustion experiment.
Anyhow, the series of the paper-combustion experiments by Timonen's group
 provides a beautiful example where a variety of theoretical achievements
 on the KPZ class can be tested in depth in experiments.

\subsubsection{Particle deposition on coffee ring.}

Very recently, Yunker \etal \cite{Yunker.etal-PRL2013}
 found an experimental realization of the KPZ-class scaling laws
 associated with the coffee ring effect.
The coffee ring effect \cite{Deegan.etal-N1997} refers to
 formation of a ring stain after evaporation of a suspension droplet
 -- the phenomenon that one finds when a droplet of coffee falls
 onto a glass table and dries out. 
This pattern results from deposition of particles driven by capillary flow
 toward the edge of the droplet \cite{Deegan.etal-N1997}.
Yunker \etal \cite{Yunker.etal-PRL2013} studied this deposition process
 with elongated particles and found three distinct regimes
 of the kinetic roughening by varying the aspect ratio of the particles.
For spheres, i.e., unit aspect ratio, they found $\beta = 0.48(4)$,
 which is explained by a spatially uncorrelated Poisson-like
 deposition process, $\beta = 1/2$.
For highly elongated particles with aspect ratios $\gtrsim 1.5$,
 they obtained $\alpha = 0.61(2)$ and $\beta = 0.68(5)$
 and compared with the quenched KPZ class.
% (see section~\ref{sec:3.2}).
In between, for the particles with aspect ratios 1.1 to 1.2,
 they revealed $\alpha = 0.51(5)$ and $\beta = 0.37(4)$,
 in agreement with the KPZ class.
Moreover, they also measured the skewness and the kurtosis
 as functions of time, which were found to converge
 to the values of the GUE Tracy-Widom distribution within the range of error.
This confirms the characteristic distribution
 for the curved-interface subclass, as we have seen in section~\ref{sec:2.2}.
From a microscopic viewpoint, particles with larger aspect ratios
 deform the air-water interface more strongly,
 leading to longer ranges of interparticle attraction
 \cite{Yunker.etal-PRL2013}.
This gives an intuitive explanation on
 why the Poisson-like and the KPZ-class growth processes arise
 for unit and intermediate aspect ratios, respectively.
% but connection to the roughening process observed for high aspect ratios
% is highly nontrivial.
Apart from the scientific importance, this experiment tells us that 
 the profound universal characteristics of the KPZ class may even underlie
 common phenomena that we experience in our daily lives.

\subsubsection{Fronts of chemical waves in disordered media}

Another very recent experiment showing the KPZ-class exponents was reported
 on fronts of chemical reactions in porous media.
Atis \etal \cite{Atis.etal-PRL2013,Atis.etal-KPZ} carried out experiments
 on the iodate arsenous acid autocatalytic reaction,
 \chem{3H_3AsO_3 + IO_3^- + 5I^- \to 3H_3AsO_4 + 6I^-},
 between two vertical acrylic glass plates,
 placed parallel to each other
 and filled with bidisperse glass beads in between.
The cell was filled with reactant solution
 and the chemical reaction was initiated from the bottom end,
 by immersing it in a reservoir of reagent containing autocatalyst.
Without external flow, the reaction forms a smooth horizontal front,
 propagating upward with a stationary concentration profile.
However, in the presence of flow
 introduced by injection or suction of the solution,
 the smooth front profile is destabilized
 by disordered flow in the porous medium, forming self-affine interfaces.
Atis \etal then found four different regimes
 by varying the flow rate:
 (1) supportive (upward) flow, downstream (upward) front propagation (SD),
 (2) weak adverse (downward) flow, upstream (upward) front propagation (AU),
 (3) moderate adverse (downward) flow, eventually static (frozen) front (AS),
 and (4) strong adverse (downward) flow,
 downstream (downward) front propagation (AD)
 \cite{Atis.etal-PRL2013}, which were also reproduced
 by their lattice Boltzmann simulations \cite{Saha.etal-EL2013}.
They experimentally measured the two scaling exponents $\alpha$ and $\beta$
 for the four regimes and found that those for the SD and AD regimes
 are consistent with the KPZ-class exponents $\alpha = 1/2$ and $\beta = 1/3$,
 while for the AS regime\footnote{
In the AS regime, front profiles are eventually frozen,
 forming a sawtoothlike pattern \cite{Atis.etal-PRL2013}.
At a given time before the complete formation of the final pattern,
 some parts of the front are already frozen, while others are still evolving.
The scaling exponents were measured
 only from these evolving parts of the front \cite{Atis.etal-KPZ}.
}
 they found $0.6 \lesssim \alpha, \beta \lesssim 0.7$,
 close to the values for the quenched KPZ class \cite{Atis.etal-KPZ}.
The appearance of the KPZ class and the quenched KPZ class
 for the strong- and weak-flow regimes, respectively, is remarkably consistent
 with the theoretical expectation drawn from the DP-type models
 for the depinning transitions (see the beginning of section~\ref{sec:3.2}).
%Indeed, the dynamics in the AS regime
% inferred from the lattice Boltzmann simulations \cite{Saha.etal-EL2013}
% is reminiscent of that of the DP depinning models
% \cite{Tang.Leschhorn-PRA1992,Buldyrev.etal-PRA1992}.
Atis \etal also described the front propagation in such disordered flow
 by the eikonal approximation,
 and found it equivalent with the quenched KPZ equation,
 further supporing their experimental findings \cite{Atis.etal-KPZ}.

\section{Remarks on analyses of scaling laws in experiments}  \label{sec:4}

As already stressed,
 absorbing-state transitions and growing interfaces
 have been extensively studied by numerical simulations,
 but not by as many experiments,
 especially if it concerns universality.
As a result, analyses used by numerical studies
 are often employed for experimental work.
Such methods are of course valid in experiments,
 but not necessarily optimal,
 because experiments usually have stronger restrictions
 on the measurable quantities
 and on the number of available independent experimental runs.
In addition, secondary or uncontrolled factors
 may intervene in the phenomenon of interest.
In view of this situation, the present section describes
 some general and practical remarks on analyses of the scaling laws
 for absorbing-state transitions and for growing interfaces,
 focusing in particular on how to test the critical behavior
 of the DP/conserved-DP class for the former,
 and the universal distributions of the $(1+1)$-dimensional KPZ class
 for the latter.

\subsection{Experimental tests of the DP/conserved-DP class}
\label{sec:4.1}

Numerical studies over many years have established
 the following three main approaches to characterize critical behavior
 of absorbing-state transitions:
 steady-state measurement, critical quenching, and critical spreading,
 as we have already seen briefly in section~\ref{sec:2.1}
 (see reviews \cite{Hinrichsen-AP2000,Henkel.etal-Book2009}
 for detailed protocols).
For the critical quenching,
 one measures relaxation of the order parameter
 from a fully active initial condition,
 which can typically be achieved by a sudden change
 in the value of the control parameter
 (the applied voltage for the electroconvection).
%This determines the critical exponent $\alpha$
% and the scaling function $F_\rho(\cdot)$ via equation~\eqref{eq:quench}.
For the critical spreading,
 one tracks the fate of a single active patch
 and measures the probability
 that the resulting active cluster survives until time $t$,
 together with its volume and squared radius.
First of all, as remarked in section~\ref{sec:3.1},
 the DP-class criticality should be tested at least by three independent
 critical exponents, hopefully more.
To this end, it is always worth trying all the above three approaches,
 whenever possible.
Data collapse as in figure~\ref{fig2}(b) is also helpful;
 comparing the constructed scaling function
 with numerical data for the DP class\footnote{
Numerical data for some of the (2+1)-dimensional DP-class scaling functions
 are available from the author upon request;
 namely, those for the order parameter relaxation in critical quenching
 (figure~\ref{fig2}(b) inset),
 as well as those for the survival probability, the volume expansion,
 and the squared radius of the active cluster in critical spreading.
} or another
 provides information complementary to the critical exponents.

That having said, those three protocols have strengths and weaknesses.
Numerically, the critical spreading protocol is widely regarded
 as the best method, since one only needs to evolve
 a few number of active sites for most of runs
 and the measurement is completely free from finite-size effects
 \cite{Hinrichsen-AP2000,Henkel.etal-Book2009}.
However, it is far less useful in experiments,
 because one does not benefit from the former point
 and will be faced with poor statistics for the survival probability.
One also needs to develop a method to introduce a single active patch
 into the given experimental system.
Concerning the steady-state measurement,
 while most experimental studies have relied on it,
 it is known to be problematic, mainly because,
 for finite-size systems, the system eventually experiences
 an unfortunate moment at which all active patches disappear at the same time.
In this sense, the system always ends up with an absorbing state,
 even if it is apparently in the active phase.
Of course, this does not occur in practice
 if the system size is large enough,
 and if the system is not too close to the critical point.
Conversely, as far as an absorbing-state transition is concerned,
 one is expected to encounter such an event near the critical point;
 otherwise spontaneous nucleation of active patches may be present
 at a nonnegligible rate, wiping out the critical behavior
 expected for absorbing-state transitions.
In contrast, the critical quench protocol seems to be
 possible and useful in many experiments.
This allows us to compare the scaling function $F_\rho(\cdot)$
 in addition to the critical exponents, as exemplified in figure~\ref{fig2}(b).
Finally, for some experiments, one is obliged to probe critical behavior
 through only a few observables, which are not necessarily appropriate
 quantities to characterize critical phenomena.
In this case, one may use hysteresis
 to extract an estimate of the exponent $\beta'$ \cite{Takeuchi-PRE2008}:
 if spontaneous nucleation rate is tiny but non-zero,
% which is usually the case in experiments,
 which is a realistic assumption for experiments,
 ramping the control parameter up and down through the critical point
 produces hysteresis loops, whose average size scales
 as $\sim r^{1/(\beta'+1)}$ with the ramping rate $r$.
Because one can choose any observable to measure hysteresis
 associated with the critical behavior
 (as far as such hysteresis is visible),
 this method may be useful in many experimental systems.

\subsection{Experimental tests of the KPZ-class universal fluctuations}
\label{sec:4.2}

Concerning the scale-invariant fluctuations of growing interfaces,
 although the methods to measure the scaling exponents $\alpha, \beta, z$
 through the Family-Vicsek scaling \eqref{eq:FamilyVicsekWidth}
 are now quite standard and have been applied to many experiments
 \cite{Barabasi.Stanley-Book1995,Meakin-PR1993,HalpinHealy.Zhang-PR1995,Krug-AP1997},
 experimental results are often more complicated, because of,
 among others, different kinds of crossover effects.
First, the Family-Vicsek scaling of course arises for lengths and times
 much larger than characteristic scales of microscopic interactions.
Even if this condition is met
 and the asymptotic growth is indeed governed by the KPZ class,
 the nonlinear term $(\nabla h)^2$ is negligible at the early times
 because the interfaces are not yet sufficiently rough;
 therefore one expects crossover from the linear growth regime,
 called the Edwards-Wilkinson regime
 ($\alpha = 1/2$ and $\beta = 1/4$ in one dimension),
 to the KPZ regime ($\alpha = 1/2, \beta = 1/3$) \cite{Krug-AP1997}.
This can take place quite late, if nonlinearity in the growth process is weak.
To add, in the presence of quenched disorder,
 as is often the case in experiments, one has additional crossover effects
 as discussed in the beginning of section~\ref{sec:3.2}.
The presence of a wall may also introduce another crossover effect
 \cite{Allegra.etal-a2013}.
One needs to overcome all these crossover effects
 to infer asymptotic scaling behavior,
 which is difficult in many experiments
 because of the rather limited ranges of accessible scales.

Despite this difficulty, we have seen in section~\ref{sec:3.2} that
 some recent experiments have clearly identified
 the scaling exponents of the $(1+1)$-dimensional KPZ class.
Then, in view of the recent analytical developments
 discussed in section~\ref{sec:2.2}, a central question to ask is
 whether the universality beyond the scaling exponents,
 at the level of the distribution and correlation functions,
 is still valid for those experimental systems.
Since only few experiments have tested this universality yet,
 here we discuss a general procedure for this,
 assuming that the KPZ-class exponents are already identified
 in a given experimental (or numerical) system.

\begin{figure}[t]
 \centering
  \includegraphics[clip,width=\hsize]{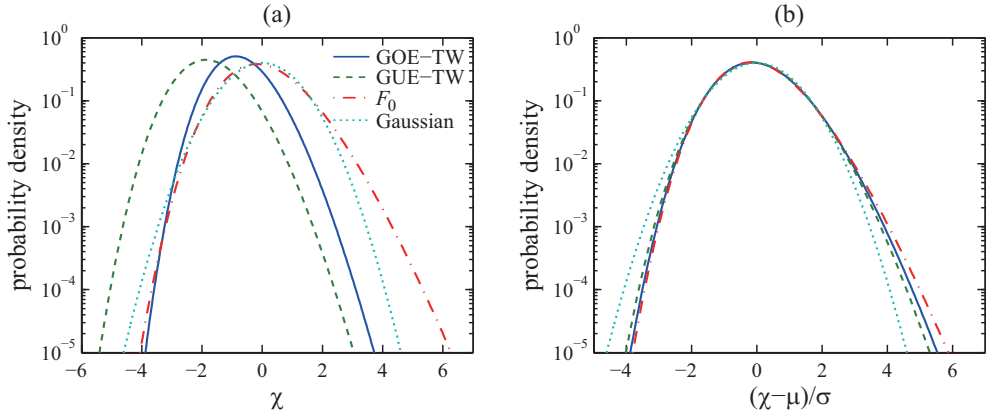}
  \caption{Comparison of the GOE and GUE Tracy-Widom distributions (blue solid and green dashed lines, respectively), the Baik-Rains $F_0$ distribution (red dashed-dotted line), and the Gaussian distribution with zero mean and unit variance (turquoise dotted line). Shown in panels (a,b) are the theoretical curves for their probability density functions and those normalized to have zero mean and unit variance, respectively ($\mu \equiv \expct{\chi}, \sigma^2 \equiv \cum{\chi^2}$). Note that the conventional definition of the random variable $\chi$ for the GOE Tracy-Widom distribution \cite{Tracy.Widom-CMP1996} is multiplied by $2^{-2/3}$ to conform with the theoretical prediction \cite{Prahofer.Spohn-PRL2000,Kriecherbauer.Krug-JPA2010,Sasamoto.Spohn-JSM2010,Corwin-RMTA2012}.}
  \label{fig6}
\end{figure}%

First of all, one should be aware that
% the difference among
 the theoretical distribution functions discussed in this context,
 namely the Tracy-Widom distributions
 and the Baik-Rains $F_0$ distribution, have only quantitative differences
 in their functional forms.
While they can readily be distinguished
 in their original definition of the stochastic variable
 (figure~\ref{fig6}(a)),
 the difference is quite subtle when they are arbitrarily shifted and scaled
 (figure~\ref{fig6}(b)).
When they are normalized to have zero mean and unit variance, for example,
 the difference is barely visible only beneath approximately $10^{-4}$
 in the probability density,
 which is usually obscured by statistical and experimental errors
 in most experiments.
Therefore, it is essential to rescale the height variable $h$ appropriately
 according to equation~\eqref{eq:Height},
 by using the two constant parameters $v_\infty$ and $\Gamma$,
 instead of fitting the theoretical distribution functions
 by adjusting the mean and the variance as free parameters.

The two parameters $v_\infty$ and $\Gamma$ can be determined as follows:
\begin{enumerate}
\item
Estimate the asymptotic growth rate $v_\infty$
 by measuring $\rd \expct{h}/\rd t$ as a function of time.
Since $\rd \expct{h}/\rd t \simeq v_\infty + at^{-2/3}$ with a constant $a$,
 one may plot $\rd \expct{h}/\rd t$ against $t^{-2/3}$
 and make a linear extrapolation to read the $y$-intercept.
\item
Determine the amplitude $\Gamma$ of the interface fluctuations.
This can be obtained by measuring the second-order cumulant
 $\expct{h^2}_{\rm c} \equiv \expct{(h-\expct{h})^2} \simeq (\Gamma t)^{2/3} \expct{\chi^2}_{\rm c}$.
Note that one needs to set here the value of $\expct{\chi^2}_{\rm c}$.
This can be chosen arbitrary, e.g.,
 $\expct{\chi^2}_{\rm c} = 1$,
 but one may choose the variance of the compared theoretical distribution
 to facilitate the comparison.
This choice does not bias the results \cite{Takeuchi.etal-SR2011}.
Alternatively, Alves \etal \cite{Alves.etal-JSM2013}
 proposed estimating $\Gamma$ from the first-order cumulant
 by $a = \Gamma^{1/3} \expct{\chi}/3$,
 on the basis of their numerical simulations.
While they found it more accurate than that measured
 from the second-order cumulant for some of the models they studied
 \cite{Alves.etal-JSM2013},
 this method did not work better for the liquid-crystal experiment.
The precision of the two estimates is determined by finite-time corrections
 in the corresponding cumulants, which are system-dependent,
 and so is the accuracy of the two methods.
Note also that the parameter $\Gamma$ may be estimated
 from coefficients of the KPZ equation
 \cite{Takeuchi.Sano-PRL2010,Takeuchi.Sano-JSP2012,Alves.etal-JSM2013},
% at least numerically,
% following procedures described in \cite{Krug.etal-PRA1992},
% but this seems difficult in typical experiments because of restrictions
% on the observation time and on the number of samples or experimental runs
% (see, however, \cite{Maunuksela.etal-EPJB2003})
 but recent numerical and theoretical work showed that
 this relation
% between $\Gamma$ and the coefficients of the KPZ equation
 may be changed in the presence of correlated noise (even short-ranged one)
 \cite{Agoritsas.etal-PRE2012},
 which is unavoidable in experimental systems.
Therefore, it would be safer to estimate $\Gamma$
% either from the first-order cumulant or from the second-order one.
 directly from the cumulants of height fluctuations, as explained here.
\end{enumerate}

Once the two parameters $v_\infty$ and $\Gamma$ are determined,
 one can make a histogram
 of the rescaled height $q \equiv (h - v_\infty t)/(\Gamma t)^{1/3}$
 to infer the distribution of the random variable $\chi$.
This can be directly compared
 with probability density functions as plotted in figure~\ref{fig6}(a).
In addition, it is always worth plotting time series of the differences
 in the $n$th-order cumulants, $\cum{q^n} - \cum{\chi_\mathrm{theory}^n}$.
This not only provides a quantitative measure of the agreement
 with the theoretical distribution, but also yields finite-time corrections
 to the asymptotic distribution, whose universal aspect
 is not fully understood yet
 \cite{Takeuchi.Sano-JSP2012,Alves.etal-JSM2013,Ferrari.Frings-JSP2011}.
% between the experimental data and the compared theoretical distributions.
Finally, using the rescaled variables,
 one can investigate many other statistical properties
 related to space-time correlation.
Experimental (or numerical) investigations of time correlation
 are particularly important, as it still remains
 unsolved by analytical means.

\section{Concluding remarks}  \label{sec:5}

Physics of scale-invariant phenomena driven far from equilibrium
 needs further investigations in many aspects,
 especially when one compares with their equilibrium counterparts.
As one cannot rely on the celebrated equilibrium statistical mechanics,
 connection between microscopic evolution and resulting macroscopic dynamics
 is a very delicate problem for out-of-equilibrium systems.
This makes it much more difficult to discriminate
 between relevant and irrelevant ingredients
% in the sense of renormalization group,
 that determine the universality class.
In this respect, experimental investigations
% of out-of-equilibrium scaling laws
 play a unique role in testing
 the robustness of the out-of-equilibrium universality under real situations,
 in the presence of nontrivial interactions,
 quenched disorder (more or less),
 self-generated out-of-equilibrium noise, and so forth.
Experiments thereby complement numerical studies,
 which can control ingredients of the system as desired,
 albeit sometimes oversimplified compared with real systems.

As overviewed in this article,
 recent years have marked considerable progress
 in experimental studies on out-of-equilibrium scaling laws
 for absorbing-state transitions and growing interfaces,
 with a growing number of experiments showing direct link
 or evidence for the DP class and the KPZ class.
% in particular regarding the DP class and the KPZ class.
%In particular, clear experimental examples
% have been found for the fundamental universality classes
% for these two situations, namely the DP class and the KPZ class,
% the latter confirming the universality
% at the level of the distribution and correlation functions.
Having more experimental examples will not only reinforce
 powerful universality of the out-of-equilibrium scaling laws;
 by comparing with other experimental systems,
 one may elucidate what are the relevant parameters that separate
 experimental systems following and not following
 the simple universality scenario in practical situations.
Let us also recall that both the DP class and the KPZ class
 correspond to the most fundamental situation
 without any additional symmetry or conservation laws.
Firm experimental grounds for these two classes would therefore help
 exploring other universality classes
 in the presence of extra symmetry or conservation
 in experimental systems,
 as partly discussed on the conserved DP class in section~\ref{sec:3.1}.
Finally, as stressed in this article,
 the $(1+1)$-dimensional KPZ class provides a unique situation
 for studies of out-of-equilibrium scaling laws,
 which allows direct comparison of exact theoretical solutions
 and quantitative experimental results,
 at the level of detailed statistical properties
 such as the distribution and correlation functions.
While theoretical studies can deal with situations
 hard to realize experimentally,
 experimentalists can measure statistical properties
 that are analytically intractable.
Of course numerical studies are also indispensable
 in many of these approaches.
The author believes that
 the current situation is the beginning of
 such cooperative developments among theory, simulations and experiments,
 which will deepen our understanding on out-of-equilibrium scaling laws
 and possibly provide novel perspectives on them.

\ack

The author is grateful to his coworkers for the experimental work
 presented in section~\ref{sec:2}, namely H. Chat\'e, M. Kuroda, M. Sano,
 T. Sasamoto and H. Spohn.
The author also thanks S. Atis, B. Hof, and L.-H. Tang
 for useful comments on the manuscript.
The author acknowledges the theoretical curves
 for the Tracy-Widom distributions provided by M. Pr\"ahofer
 (figures~\ref{fig3}(d) and \ref{fig6}),
 that for the Baik-Rains $F_0$ distribution by T. Imamura
 (figure~\ref{fig6}),
 and those for the Airy$_1$ and Airy$_2$ correlation functions
 by F. Bornemann (figure~\ref{fig4}(a)), the last ones being
 obtained by an algorithm described in \cite{Bornemann-MC2010}.

\section*{References}

\bibliographystyle{iopart-num}
\bibliography{statphys}

\providecommand{\newblock}{}
\begin{thebibliography}{100}
\expandafter\ifx\csname url\endcsname\relax
  \def\url#1{{\tt #1}}\fi
\expandafter\ifx\csname urlprefix\endcsname\relax\def\urlprefix{URL }\fi
\providecommand{\eprint}[2][]{\url{#2}}

\bibitem{Goldenfeld-Book1992}
Goldenfeld N 1992 {\em Lectures on Phase Transitions and the Renormalization
  Group\/} Frontiers in Physics (Colorado: Westview Press)

\bibitem{Henkel-Book1999}
Henkel M 1999 {\em Conformal Invariance and Critical Phenomena\/} (Berlin:
  Springer)

\bibitem{Wilson-RMP1975}
Wilson K~G 1975 {\em Rev. Mod. Phys.\/} {\bf 47} 773--840

\bibitem{Tauber-Book2013}
T\"auber U~C (to be published) {\em Critical Dynamics: A Field Theory Approach
  to Equilibrium and Non-Equilibrium Scaling Behavior\/} (Cambridge: Cambridge
  Univ. Press)

\bibitem{Hinrichsen-AP2000}
Hinrichsen H 2000 {\em Adv. Phys.\/} {\bf 49} 815--958

\bibitem{Henkel.etal-Book2009}
Henkel M, Hinrichsen H and L\"ubeck S 2009 {\em Non-Equilibrium Phase
  Transitions: Volume 1: Absorbing Phase Transitions\/} 1st ed Theoretical and
  Mathematical Physics (Dordrecht: Springer)

\bibitem{Barabasi.Stanley-Book1995}
Barab\'asi A~L and Stanley H~E 1995 {\em Fractal Concepts in Surface Growth\/}
  (Cambridge: Cambridge Univ. Press)

\bibitem{Kardar.etal-PRL1986}
Kardar M, Parisi G and Zhang Y~C 1986 {\em Phys. Rev. Lett.\/} {\bf 56}
  889--892

\bibitem{Meakin-PR1993}
Meakin P 1993 {\em Phys. Rep.\/} {\bf 235} 189--289

\bibitem{HalpinHealy.Zhang-PR1995}
Halpin-Healy T and Zhang Y~C 1995 {\em Phys. Rep.\/} {\bf 254} 215--414

\bibitem{Krug-AP1997}
Krug J 1997 {\em Adv. Phys.\/} {\bf 46} 139--282

\bibitem{Ciliberto.Bigazzi-PRL1988}
Ciliberto S and Bigazzi P 1988 {\em Phys. Rev. Lett.\/} {\bf 60} 286--289

\bibitem{Daviaud.etal-PRA1990}
Daviaud F, Bonetti M and Dubois M 1990 {\em Phys. Rev. A\/} {\bf 42} 3388--3399

\bibitem{Takeuchi.etal-PRL2007}
Takeuchi K~A, Kuroda M, Chat\'e H and Sano M 2007 {\em Phys. Rev. Lett.\/} {\bf
  99} 234503

\bibitem{Takeuchi.etal-PRE2009}
Takeuchi K~A, Kuroda M, Chat\'e H and Sano M 2009 {\em Phys. Rev. E\/} {\bf 80}
  051116

\bibitem{Takeuchi.Sano-PRL2010}
Takeuchi K~A and Sano M 2010 {\em Phys. Rev. Lett.\/} {\bf 104} 230601

\bibitem{Takeuchi.etal-SR2011}
Takeuchi K~A, Sano M, Sasamoto T and Spohn H 2011 {\em Sci. Rep.\/} {\bf 1} 34

\bibitem{Takeuchi.Sano-JSP2012}
Takeuchi K~A and Sano M 2012 {\em J. Stat. Phys.\/} {\bf 147} 853--890

\bibitem{deGennes.Prost-Book1995}
de~Gennes P~G and Prost J 1995 {\em The Physics of Liquid Crystals\/} 2nd ed
  ({\em International Series of Monographs on Physics\/} vol~83) (New York:
  Oxford Univ. Press)

\bibitem{Kai.Zimmermann-PTPS1989}
Kai S and Zimmermann W 1989 {\em Prog. Theor. Phys. Suppl.\/} {\bf 99} 458--492

\bibitem{Frisch-Book1995}
Frisch U 1995 {\em Turbulence: The Legacy of A. N. Kolmogorov\/} (Cambridge:
  Cambridge Univ. Press)

\bibitem{Kai.etal-JPSJ1989}
Kai S, Zimmermann W, Andoh M and Chizumi N 1989 {\em J. Phys. Soc. Jpn.\/} {\bf
  58} 3449--3452

\bibitem{Kai.etal-PRL1990}
Kai S, Zimmermann W, Andoh M and Chizumi N 1990 {\em Phys. Rev. Lett.\/} {\bf
  64} 1111--1114

\bibitem{Fazio.Komitov-EL1999}
Fazio V~S~U and Komitov L 1999 {\em Europhys. Lett.\/} {\bf 46} 38--42

\bibitem{Grassberger.Zhang-PA1996}
Grassberger P and Zhang Y~C 1996 {\em Physica A\/} {\bf 224} 169--179

\bibitem{Voigt.Ziff-PRE1997}
Voigt C~A and Ziff R~M 1997 {\em Phys. Rev. E\/} {\bf 56} R6241--R6244

\bibitem{Pomeau-PD1986}
Pomeau Y 1986 {\em Physica D\/} {\bf 23} 3--11

\bibitem{Janssen-ZPB1981}
Janssen H~K 1981 {\em Z. Phys. B\/} {\bf 42} 151--154

\bibitem{Grassberger-ZPB1982}
Grassberger P 1982 {\em Z. Phys. B\/} {\bf 47} 365--374

\bibitem{Family.Vicsek-JPA1985}
Family F and Vicsek T 1985 {\em J. Phys. A\/} {\bf 18} L75--L81

\bibitem{Mehta-Book2004}
Mehta M~L 2004 {\em Random Matrices\/} 3rd ed ({\em Pure and Applied
  Mathematics\/} vol 142) (San Diego: Elsevier)

\bibitem{Tracy.Widom-CMP1994}
Tracy C~A and Widom H 1994 {\em Commun. Math. Phys.\/} {\bf 159} 151--174

\bibitem{Tracy.Widom-CMP1996}
Tracy C~A and Widom H 1996 {\em Commun. Math. Phys.\/} {\bf 177} 727--754

\bibitem{Prahofer.Spohn-PRL2000}
Pr\"ahofer M and Spohn H 2000 {\em Phys. Rev. Lett.\/} {\bf 84} 4882--4885

\bibitem{Kriecherbauer.Krug-JPA2010}
Kriecherbauer T and Krug J 2010 {\em J. Phys. A\/} {\bf 43} 403001

\bibitem{Sasamoto.Spohn-JSM2010}
Sasamoto T and Spohn H 2010 {\em J. Stat. Mech.\/} {\bf 2010} P11013

\bibitem{Corwin-RMTA2012}
Corwin I 2012 {\em Random Matrices Theory Appl.\/} {\bf 1} 1130001

\bibitem{Johansson-CMP2000}
Johansson K 2000 {\em Commun. Math. Phys.\/} {\bf 209} 437--476

\bibitem{Sasamoto.Spohn-PRL2010}
Sasamoto T and Spohn H 2010 {\em Phys. Rev. Lett.\/} {\bf 104} 230602

\bibitem{Amir.etal-CPAM2011}
Amir G, Corwin I and Quastel J 2011 {\em Commun. Pure Appl. Math.\/} {\bf 64}
  466--537

\bibitem{Calabrese.LeDoussal-PRL2011}
Calabrese P and Le~Doussal P 2011 {\em Phys. Rev. Lett.\/} {\bf 106} 250603

\bibitem{Bornemann.etal-JSP2008}
Bornemann F, Ferrari P~L and Pr\"ahofer M 2008 {\em J. Stat. Phys.\/} {\bf 133}
  405--415

\bibitem{Takeuchi-JSM2012}
Takeuchi K~A 2012 {\em J. Stat. Mech.\/} {\bf 2012} P05007

\bibitem{Singha-JSM2005}
Singha S~B 2005 {\em J. Stat. Mech.\/} {\bf 2005} P08006

\bibitem{Baik.Rains-JSP2000}
Baik J and Rains E~M 2000 {\em J. Stat. Phys.\/} {\bf 100} 523--541

\bibitem{Takeuchi-PRL2013}
Takeuchi K~A 2013 {\em Phys. Rev. Lett.\/} {\bf 110} 210604

\bibitem{Buldyrev.etal-PRA1992}
Buldyrev S~V, Barab\'asi A~L, Caserta F, Havlin S, Stanley H~E and Vicsek T
  1992 {\em Phys. Rev. A\/} {\bf 45} R8313--R8316

\bibitem{Michalland.etal-EL1993}
Michalland S, Rabaud M and Couder Y 1993 {\em Europhys. Lett.\/} {\bf 22}
  17--22

\bibitem{Willaime.etal-PRE1993}
Willaime H, Cardoso O and Tabeling P 1993 {\em Phys. Rev. E\/} {\bf 48}
  288--295

\bibitem{Degen.etal-PRE1996}
Degen M~M, Mutabazi I and David~Andereck C 1996 {\em Phys. Rev. E\/} {\bf 53}
  3495--3504

\bibitem{Colovas.Andereck-PRE1997}
Colovas P~W and Andereck C~D 1997 {\em Phys. Rev. E\/} {\bf 55} 2736--2741

\bibitem{Daerr.Douady-N1999}
Daerr A and Douady S 1999 {\em Nature\/} {\bf 399} 241--243

\bibitem{Hinrichsen.etal-PRL1999}
Hinrichsen H, Jim\'enez-Dalmaroni A, Rozov Y and Domany E 1999 {\em Phys. Rev.
  Lett.\/} {\bf 83} 4999--5002

\bibitem{Cros.LeGal-PF2002}
Cros A and Le~Gal P 2002 {\em Phys. Fluids\/} {\bf 14} 3755--3765

\bibitem{Rupp.etal-PRE2003}
Rupp P, Richter R and Rehberg I 2003 {\em Phys. Rev. E\/} {\bf 67} 036209

\bibitem{Lepiller.etal-PF2007}
Lepiller V, Prigent A, Dumouchel F and Mutabazi I 2007 {\em Phys. Fluids\/}
  {\bf 19} 054101

\bibitem{Latrache.etal-PRE2012}
Latrache N, Crumeyrolle O and Mutabazi I 2012 {\em Phys. Rev. E\/} {\bf 86}
  056305

\bibitem{Jensen-JPA1999}
Jensen I 1999 {\em J. Phys. A\/} {\bf 32} 5233--5249

\bibitem{Pirat.etal-PRL2005}
Pirat C, Naso A, Meunier J~L, Ma\"\i{}ssa P and Mathis C 2005 {\em Phys. Rev.
  Lett.\/} {\bf 94} 134502

\bibitem{Reynolds-PTRSL1883}
Reynolds O 1883 {\em Phil. Trans. R. Soc. London\/} {\bf 174} 935--982

\bibitem{Wygnanski.Champagne-JFM1973}
Wygnanski I~J and Champagne F~H 1973 {\em J. Fluid. Mech.\/} {\bf 59} 281--335

\bibitem{Hof.etal-S2004}
Hof B, van Doorne C~W~H, Westerweel J, Nieuwstadt F~T~M, Faisst H, Eckhardt B,
  Wedin H, Kerswell R~R and Waleffe F 2004 {\em Science\/} {\bf 305} 1594--1598

\bibitem{Hof.etal-N2006}
Hof B, Westerweel J, Schneider T~M and Eckhardt B 2006 {\em Nature\/} {\bf 443}
  59--62

\bibitem{Hof.etal-PRL2008}
Hof B, de~Lozar A, Kuik D~J and Westerweel J 2008 {\em Phys. Rev. Lett.\/} {\bf
  101} 214501

\bibitem{Avila.etal-S2011}
Avila K, Moxey D, de~Lozar A, Avila M, Barkley D and Hof B 2011 {\em Science\/}
  {\bf 333} 192--196

\bibitem{Kuik.etal-JFM2010}
Kuik D~J, Poelma C and Westerweel J 2010 {\em J. Fluid. Mech.\/} {\bf 645}
  529--539

\bibitem{Avila.etal-JFM2010}
Avila M, Willis A~P and Hof B 2010 {\em J. Fluid. Mech.\/} {\bf 646} 127--136

\bibitem{Samanta.etal-JFM2011}
Samanta D, Lozar A~D and Hof B 2011 {\em J. Fluid. Mech.\/} {\bf 681} 193--204

\bibitem{Avila.Hof-PRE2013}
Avila M and Hof B 2013 {\em Phys. Rev. E\/} {\bf 87} 063012

\bibitem{Moxey.Barkley-PNAS2010}
Moxey D and Barkley D 2010 {\em Proc. Natl. Acad. Sci. USA\/} {\bf 107}
  8091--8096

\bibitem{Shi.etal-PRL2013}
Shi L, Avila M and Hof B 2013 {\em Phys. Rev. Lett.\/} {\bf 110} 204502

\bibitem{Pine.etal-N2005}
Pine D~J, Gollub J~P, Brady J~F and Leshansky A~M 2005 {\em Nature\/} {\bf 438}
  997--1000

\bibitem{Corte.etal-NP2008}
Cort\'e L, Chaikin P~M, Gollub J~P and Pine D~J 2008 {\em Nat. Phys.\/} {\bf 4}
  420--424

\bibitem{Lubeck-IJMPB2004}
L\"ubeck S 2004 {\em Int. J. Mod. Phys. B\/} {\bf 18} 3977--4118

\bibitem{Menon.Ramaswamy-PRE2009}
Menon G~I and Ramaswamy S 2009 {\em Phys. Rev. E\/} {\bf 79} 061108

\bibitem{Franceschini.etal-PRL2011}
Franceschini A, Filippidi E, Guazzelli E and Pine D~J 2011 {\em Phys. Rev.
  Lett.\/} {\bf 107} 250603

\bibitem{Mangan.etal-PRL2008}
Mangan N, Reichhardt C and Reichhardt C~J~O 2008 {\em Phys. Rev. Lett.\/} {\bf
  100} 187002

\bibitem{Reichhardt.Reichhardt-PRL2009}
Reichhardt C and Reichhardt C~J~O 2009 {\em Phys. Rev. Lett.\/} {\bf 103}
  168301

\bibitem{Okuma.etal-PRB2011}
Okuma S, Tsugawa Y and Motohashi A 2011 {\em Phys. Rev. B\/} {\bf 83} 012503

\bibitem{Okuma.etal-JPSJ2012}
Okuma S, Kawamura Y and Tsugawa Y 2012 {\em J. Phys. Soc. Jpn.\/} {\bf 81}
  114718

\bibitem{Basu.etal-PRL2012}
Basu M, Basu U, Bondyopadhyay S, Mohanty P~K and Hinrichsen H 2012 {\em Phys.
  Rev. Lett.\/} {\bf 109} 015702

\bibitem{Basu.etal-EPJB2013}
Basu U, Basu M and Mohanty P~K 2013 {\em Eur. Phys. J. B\/} {\bf 86} 1--7

\bibitem{Vespignani.etal-PRL1998}
Vespignani A, Dickman R, Mu\~noz M~A and Zapperi S 1998 {\em Phys. Rev.
  Lett.\/} {\bf 81} 5676--5679

\bibitem{Bonachela.Munoz-PA2007}
Bonachela J~A and Mu\~noz M~A 2007 {\em Physica A\/} {\bf 384} 89--93

\bibitem{Alava.etal-AP2004}
Alava M, Dub\'e M and Rost M 2004 {\em Adv. Phys.\/} {\bf 53} 83--175

\bibitem{Tang.Leschhorn-PRA1992}
Tang L~H and Leschhorn H 1992 {\em Phys. Rev. A\/} {\bf 45} R8309--R8312

\bibitem{Amaral.etal-PRL1994}
Amaral L~A~N, Barab\'asi A~L and Stanley H~E 1994 {\em Phys. Rev. Lett.\/} {\bf
  73} 62--65

\bibitem{Amaral.etal-PRE1995}
Amaral L~A~N, Barab\'asi A~L, Buldyrev S~V, Harrington S~T, Havlin S,
  Sadr-Lahijany R and Stanley H~E 1995 {\em Phys. Rev. E\/} {\bf 51} 4655--4673

\bibitem{Tang.etal-PRL1995}
Tang L~H, Kardar M and Dhar D 1995 {\em Phys. Rev. Lett.\/} {\bf 74} 920--923

\bibitem{Leschhorn.etal-APB1997}
Leschhorn H, Nattermann T, Stepanow S and Tang L~H 1997 {\em Ann. Phys.
  (Berlin)\/} {\bf 509} 1--34

\bibitem{Csahok.etal-JPA1993}
Csah\'ok Z, Honda K and Vicsek T 1993 {\em J. Phys. A\/} {\bf 26} L171--L178

\bibitem{Csahok.etal-PA1993}
Csah\'ok Z, Honda K, Somfai E, Vicsek M and Vicsek T 1993 {\em Physica A\/}
  {\bf 200} 136--154

\bibitem{Medina.etal-PRA1989}
Medina E, Hwa T, Kardar M and Zhang Y~C 1989 {\em Phys. Rev. A\/} {\bf 39}
  3053--3075

\bibitem{Zhang-JP1990}
Zhang Y~C 1990 {\em J. Phys. (Paris)\/} {\bf 51} 2129--2134

\bibitem{Horvath.etal-PRL1991}
Horv\'ath V~K, Family F and Vicsek T 1991 {\em Phys. Rev. Lett.\/} {\bf 67}
  3207--3210

\bibitem{Kertesz.etal-F1993}
Kert\'esz J, Horv\'ath V~K and Weber F 1993 {\em Fractals\/} {\bf 1} 67--74

\bibitem{Engoy.etal-PRL1994}
Eng\o{}y T, M\aa{}l\o{}y K~J, Hansen A and Roux S 1994 {\em Phys. Rev. Lett.\/}
  {\bf 73} 834--837

\bibitem{Degawa.etal-PRL2006}
Degawa M, Stasevich T~J, Cullen W~G, Pimpinelli A, Einstein T~L and Williams
  E~D 2006 {\em Phys. Rev. Lett.\/} {\bf 97} 080601

\bibitem{Matsushita.etal-B2004}
Matsushita M, Hiramatsu F, Kobayashi N, Ozawa T, Yamazaki Y and Matsuyama T
  2004 {\em Biofilms\/} {\bf 1} 305--317

\bibitem{Vicsek.etal-PA1990}
Vicsek T, Cserz\H{o} M and Horv\'ath V~K 1990 {\em Physica A\/} {\bf 167}
  315--321

\bibitem{Wakita.etal-JPSJ1997}
Wakita J~i, Itoh H, Matsuyama T and Matsushita M 1997 {\em J. Phys. Soc.
  Jpn.\/} {\bf 66} 67--72

\bibitem{Hallatschek.etal-PNAS2007}
Hallatschek O, Hersen P, Ramanathan S and Nelson D~R 2007 {\em Proc. Natl.
  Acad. Sci. USA\/} {\bf 104} 19926--19930

\bibitem{Huergo.etal-PRE2010}
Huergo M~A~C, Pasquale M~A, Bolz\'an A~E, Arvia A~J and Gonz\'alez P~H 2010
  {\em Phys. Rev. E\/} {\bf 82} 031903

\bibitem{Huergo.etal-PRE2011}
Huergo M~A~C, Pasquale M~A, Gonz\'alez P~H, Bolz\'an A~E and Arvia A~J 2011
  {\em Phys. Rev. E\/} {\bf 84} 021917

\bibitem{Huergo.etal-PRE2012}
Huergo M~A~C, Pasquale M~A, Gonz\'alez P~H, Bolz\'an A~E and Arvia A~J 2012
  {\em Phys. Rev. E\/} {\bf 85} 011918

\bibitem{Zhang.etal-PA1992}
Zhang J, Zhang Y~C, Alstr\o{}m P and Levinsen M~T 1992 {\em Physica A\/} {\bf
  189} 383--389

\bibitem{Maunuksela.etal-PRL1997}
Maunuksela J, Myllys M, K\"ahk\"onen O~P, Timonen J, Provatas N, Alava M~J and
  Ala-Nissila T 1997 {\em Phys. Rev. Lett.\/} {\bf 79} 1515--1518

\bibitem{Myllys.etal-PRE2001}
Myllys M, Maunuksela J, Alava M, Ala-Nissila T, Merikoski J and Timonen J 2001
  {\em Phys. Rev. E\/} {\bf 64} 036101

\bibitem{Maunuksela.etal-EPJB2003}
Maunuksela J, Myllys M, Merikoski J, Timonen J, K\"arkk\"ainen T, Welling M and
  Wijngaarden R 2003 {\em Eur. Phys. J. B\/} {\bf 33} 193--202

\bibitem{Merikoski.etal-PRL2003}
Merikoski J, Maunuksela J, Myllys M, Timonen J and Alava M~J 2003 {\em Phys.
  Rev. Lett.\/} {\bf 90} 024501

\bibitem{Kallabis.Krug-EL1999}
Kallabis H and Krug J 1999 {\em Europhys. Lett.\/} {\bf 45} 20--25

\bibitem{Ferrari.Frings-JSM2013}
Ferrari P~L and Frings R 2013 {\em J. Stat. Mech.\/} {\bf 2013} P02001

\bibitem{Miettinen.etal-EPJB2005}
Miettinen L, Myllys M, Merikoski J and Timonen J 2005 {\em Eur. Phys. J. B\/}
  {\bf 46} 55--60

\bibitem{Yunker.etal-PRL2013}
Yunker P~J, Lohr M~A, Still T, Borodin A, Durian D~J and Yodh A~G 2013 {\em
  Phys. Rev. Lett.\/} {\bf 110} 035501

\bibitem{Deegan.etal-N1997}
Deegan R~D, Bakajin O, Dupont T~F, Huber G, Nagel S~R and Witten T~A 1997 {\em
  Nature\/} {\bf 389} 827--829

\bibitem{Atis.etal-PRL2013}
Atis S, Saha S, Auradou H, Salin D and Talon L 2013 {\em Phys. Rev. Lett.\/}
  {\bf 110} 148301

\bibitem{Atis.etal-KPZ}
Atis S, Awadhesh K, Talon L and Salin D (in preparation)

\bibitem{Saha.etal-EL2013}
Saha S, Atis S, Salin D and Talon L 2013 {\em Europhys. Lett.\/} {\bf 101}
  38003

\bibitem{Takeuchi-PRE2008}
Takeuchi K~A 2008 {\em Phys. Rev. E\/} {\bf 77} 030103(R)

\bibitem{Allegra.etal-a2013}
Allegra N, Fortin J~Y and Henkel M 2013 {\em arXiv\/}  1309.1634

\bibitem{Alves.etal-JSM2013}
Alves S~G, Oliveira T~J and Ferreira S~C 2013 {\em J. Stat. Mech.\/} {\bf 2013}
  P05007

\bibitem{Agoritsas.etal-PRE2012}
Agoritsas E, Bustingorry S, Lecomte V, Schehr G and Giamarchi T 2012 {\em Phys.
  Rev. E\/} {\bf 86} 031144

\bibitem{Ferrari.Frings-JSP2011}
Ferrari P~L and Frings R 2011 {\em J. Stat. Phys.\/} {\bf 144} 1123--1150

\bibitem{Bornemann-MC2010}
Bornemann F 2010 {\em Math. Comput.\/} {\bf 79} 871--915

\end{thebibliography}

\end{document}